\documentclass[fleqn,usenatbib]{mnras}

\usepackage[T1]{fontenc}

\DeclareRobustCommand{\VAN}[3]{#2}
\let\VANthebibliography\thebibliography
\def\thebibliography{\DeclareRobustCommand{\VAN}[3]{##3}\VANthebibliography}

\usepackage{graphicx}	
\usepackage{amsmath}	
\usepackage{amssymb}	
\usepackage{multicol}
\usepackage{caption}
\usepackage{subcaption}
\usepackage{xcolor}
\usepackage{hyperref}
\usepackage{footnote}
\title[Rotational modulation in A and F stars]{Rotational modulation in A and F stars: Magnetic stellar spots or convective core rotation?}

\author[A. I. Henriksen et al.]{Andreea I. Henriksen,$^{1}$\thanks{E-mail: andreea@space.dtu.dk}
Victoria Antoci,$^{1}$\thanks{E-mail: antoci@space.dtu.dk}
Hideyuki Saio,$^{2}$
Matteo Cantiello,$^{3}$
Hans Kjeldsen,$^{4}$
\newauthor
Donald W. Kurtz,$^{5,6}$
Simon J. Murphy,$^{7}$
Savita Mathur,$^{8,9}$
Rafael A. Garc\'ia,$^{10}$
\^Angela R. G. Santos $^{11}$
\\
$^{1}$National Space Institute, Technical University of Denmark, Elektrovej, DK-2800 Kgs. Lyngby, Denmark\\
$^{2}$Astronomical Institute, Graduate School of Science, Tohoku University, Sendai 980-8578, Japan\\
$^{3}$Center for Computational Astrophysics, Flatiron Institute, 162 5th Avenue, New York, NY 10010, USA\\
$^{4}$Stellar Astrophysics Centre, Department of Physics and Astronomy, Aarhus University, 8000 Aarhus C, Denmark\\
$^{5}$Department of Physics, North-West University, Dr Albert Luthuli Drive, Mahikeng 2735, South Africa\\
$^{6}$Jeremiah Horrocks Institute, University of Central Lancashire, Preston PR1 2HE, UK \\
$^{7}$University of Southern Queensland - Springfield Campus: Springfield, QLD, AU\\
$^{8}$Instituto de Astrof\'isica de Canarias (IAC), E-38205 La Laguna, Tenerife, Spain\\
$^{9}$Universidad de La Laguna (ULL), Departamento de Astrof\'isica, E-38206 La Laguna, Tenerife, Spain\\
$^{10}$AIM, CEA, CNRS, Universit\'e Paris-Saclay, Universit\'e Paris Diderot, Sorbonne Paris Cit\'e, F-91191 Gif-sur-Yvette, France\\
$^{11}$ Instituto de Astrof\'isica e Ci\^encias do Espa\c{c}o, Universidade do Porto, CAUP, Rua das Estrelas, PT4150-762 Porto, Portugal
}

\date{Accepted XXX. Received YYY; in original form ZZZ}

\pubyear{20xx}
\usepackage{newtxtext,newtxmath}
\begin{document}

\defcitealias{Cantiello_2019}{CB19}
\label{firstpage}
\pagerange{\pageref{firstpage}--\pageref{lastpage}}
\maketitle

\begin{abstract}
The Kepler mission revealed a plethora of stellar variability in the light curves of many stars, some associated with magnetic activity or stellar oscillations. In this work, we analyse the periodic signal in 162 intermediate-mass stars, interpreted as Rossby modes and rotational modulation - the so-called \textit{hump \& spike} feature. We investigate whether the rotational modulation (\textit{spike}) is due to stellar spots caused by magnetic fields or due to Overstable Convective (OsC) modes resonantly exciting g~modes, with frequencies corresponding to the convective core rotation rate.
Assuming that the spikes are created by magnetic spots at the stellar surface, we recover the amplitudes of the magnetic fields, which are in good agreement with theoretical predictions.
Our data show a clear anti-correlation between the spike amplitudes and stellar mass and possibly a correlation with stellar age, consistent with the dynamo-generated magnetic fields theory in (sub)-surface convective layers. Investigating the harmonic behaviour, we find that for 125 stars neither of the two possible explanations can be excluded. While our results suggest that the dynamo-generated magnetic field scenario is more likely to explain the \textit{spike} feature, we assess further work is needed to distinguish between the two scenarios. One method for ruling out one of the two explanations is to directly observe magnetic fields in \textit{hump \& spike} stars. Another would be to impose additional constraints through detailed modelling of our stars, regarding the rotation requirement in the OsC mode scenario or the presence of a convective-core (stellar age).

\end{abstract}

\begin{keywords}
stars: early-type --
stars: rotation --
stars: magnetic fields --
stars: oscillations

\end{keywords}

\section{Introduction}

Around 10 per cent of A-type stars have been discovered to be chemically and magnetically peculiar. It is widely accepted that the magnetic fields in Ap stars are of fossil origin, i.e. they were formed in the pre-main sequence stage of stellar evolution (e.g., \citealt{1945MNRAS.105..166C, 2004Natur.431..819B, 2013MNRAS.428.2789B}). A fossil magnetic field means that at the current evolutionary stage, the star cannot maintain the magnetic field against spontaneous decay \citep{1982ARA&A..20..191B}.
The predicted strengths of fossil-magnetic fields, obtained by \citet{2004Natur.431..819B} through numerical simulation, were of the order of 10 kG, agreeing with observations. 

Magnetic fields in chemically peculiar stars inhibit the motions of ions (which have an electric charge) and force particular elements to be concentrated in spots, as many observations confirm (e.g., \citealt{2009ssc..book.....G}). As most Ap stars are slow rotators and have an overabundance of certain elements (e.g., \citealt{2006A&A...450..763K}), their spectral lines are very sharp, which allows one to directly detect the magnetic fields through magnetic splitting of the lines, or by spectro-polarimetric observations measuring the Zeeman effect (see, e.g., \citealt{2009ARA&A..47..333D}). However, our understanding of how magnetic fields are generated and operate in Ap stars cannot be extended to other types of stars that exhibit magnetic activity. 

Magnetic fields have been investigated and measured in other A stars, however most of them have weak magnetic field strengths (see e.g. \citealt{2007A&A...475.1053A}, \citealt{2009A&A...500L..41L}, \citealt{2011A&A...532L..13P}, \citealt{2016MNRAS.459L..81B, 2016A&A...586A..97B}, \citealt{2017MNRAS.468L..46N}, \citealt{ 2020MNRAS.494.5682S}). Interestingly, \citet{2007A&A...475.1053A} found a lack of stellar magnetic fields with strengths between 300 and a few Gauss, which is now commonly termed as the `magnetic desert'. The strong magnetic fields of Ap stars highlights their distinctive character with respect to non-peculiar A stars. 

\subsection{Origin of magnetic fields and magnetic activity in non-peculiar intermediate-mass stars}

Our initial understanding of stellar magnetic fields is based on studies of the Sun. The kinetic energy generated by the motions in the convective layers, coupled with differential solar rotation, is converted by the solar dynamo to magnetic energy. 
Dynamo generated magnetic fields in the Sun can then be observed at the photosphere (see, e.g., the review by \citealt{Berdyugina2005} and references therein).

The magnetic fields at the surface of the Sun can partially inhibit the convective energy transport. Such regions appear as spots, visible because they are cooler and therefore darker than the bright photosphere \citep{2009A&ARv..17..251S}. It is known from both theory and helioseismic observations that the convective zone (CZ) of the Sun is approximately 29 per cent of its outer radius (e.g. \citet{1993PhR...230...57T, 2002RvMP...74.1073C}). The solar CZ is efficient and deep enough to sustain a dynamo that can generate a magnetic field, penetrating the solar photosphere and creating sunspots. Given the proximity of the Sun, which allows us to resolve its disk, sunspots have been visually observed on its surface for centuries. In order to understand stellar and solar dynamo, however, one has to use magnetic activity observations of other stars, which expands the range of stellar parameters on which the dynamo theories can be tested \citep{Berdyugina2005,Brandenburg:2005}. Magnetic activity in late-type stars, which possess deep convective envelopes, has been observed abundantly (see \citealt[][chapter 2]{Berdyugina2005}). 

In the case of more massive early-type stars, detecting and understanding the origin of their magnetic fields present additional challenges. Nevertheless, one can infer information regarding magnetic activity by looking for rotational modulation in the photometric light curves of stars. The brightness variations can be attributed to temperature variations caused by stellar spots co-rotating with the stellar surface. While the more massive HgMn stars show spots but no surface magnetic fields have been recorded yet \citep{2013A&A...554A..61K}, stellar spots are generally accepted to be a consequence of magnetic activity also in intermediate-mass stars \citep{Berdyugina2005}. 

As the stellar structure changes with increasing stellar mass, one cannot assume that the mechanism that generates magnetic fields in higher mass stars is the same as in the Sun. This also reinforces the necessity of studying magnetic fields in more massive stars.

A debatable aspect regarding the presence of magnetic fields in non-chemically peculiar main-sequence higher-mass stars ($\geq 1.3$\,M$_\odot$) comes from the fact that, as opposed to our Sun's radiative core and deep convective envelope, the core of more massive stars is convective and surrounded by a deep radiative envelope \footnote[1]{Depending on the mass, stars with masses lower than $1.3$\,M$_\odot$, could develop a convective core during later stages of the main-sequence lifetime}. A more detailed analysis of the outermost layers of intermediate-mass stars reveals, however, that due to high opacity and/or low adiabatic gradient (conditions fulfilled in the ionization zones of H, He and/or Fe), there are thin convective layers at or below the stellar surface (e.g. \citealt{Cantiello_2019}, hereafter \citetalias{Cantiello_2019}).

\citet{2011A&A...534A.140C,Cantiello_2019}
proposed that despite their small aspect ratio, these convective regions could sustain a dynamo and generate a magnetic field. Since early-type stars tend to be rapidly-rotating, magnetic structures could also grow to scales larger than the scale of convective motions. The rapid rotation could distinguish stars with this type of dynamo-generated magnetic field from those in the slowly rotating Ap stars, which, as mentioned above, are believed to have magnetic fields generated prior to the main-sequence stage. \citetalias{Cantiello_2019} also suggest that dynamo-generated fields have rapidly evolving features of small-scale and that the strength of surface magnetic fields is expected to increase as the stars evolve. While it is difficult to predict the exact size of the magnetic features, it is expected that their sizes are correlated to the pressure scale height at the bottom of the shallow convective envelope, which is expected to be less than $1$ per cent of the stellar radius. \citetalias{Cantiello_2019} predicted surface magnetic fields strengths to reach 1kG for the lower mass stars and $\lesssim\!1$ G for stars with $T_{\rm eff}$ >10,000\,K, noting that these could be just lower limits.

Both theory and observations suggest that the transition between stars with convective surfaces to stars with radiative surfaces is at around 10,000\,K (\citetalias{Cantiello_2019}). Cool intermediate-mass stars ($T_{\rm eff} \lesssim\!10^4$\,K) likely have a very thin convective surface layer due to the ionization of H and He I (see Figure 2 in \citetalias{Cantiello_2019}). The transported flux in these convective layers, which separate the radiative envelope from the surface, is negligible because they contain very little mass.  According to \citetalias{Cantiello_2019}, magnetic fields can reach the stellar surface through magnetic buoyancy, locally changing the gas pressure and lowering the optical depth of the photosphere. Consequently, spots will appear brighter than the surrounding photosphere because the flux originates from deeper radiative areas in the star, where the temperature is higher than at the surface. 

In contrast to the dynamo-generated magnetic fields that \citetalias{Cantiello_2019} described, \citet{2013MNRAS.428.2789B} proposed the failed-fossil scenario for the origin of ultra-weak magnetic fields in A stars. The failed-fossil magnetic fields are expected to have large-scale, slowly evolving features that would decrease as the stars evolve. The predictions in \citet{2013MNRAS.428.2789B} suggest that: (1) the time scale of the failed-fossil field's evolution is around the stellar age, (2) the length scale of surface magnetic structures should be at least $20$ per cent of the radius of the star, (3) the higher stellar rotation, the stronger the stellar magnetic field. With respect to the large-scale and slowly evolving features, the characteristics of failed-fossil magnetic fields are opposite to those of dynamo-generated magnetic fields predicted by \citetalias{Cantiello_2019}, but for both scenarios the rotation should play an important role. One could test both scenarios by performing an ensemble analysis on a homogeneous sample of stars that show magnetic activity signs.

For this study, we use the, so-called, \textit{hump \& spike} stars, which 
 \citet{2018MNRAS.474.2774S} suggested could show signs of magnetic activity in the form of spots. The common characteristic of these stars is a feature in the Fourier spectrum (see Fig.\,\ref{fig:zoo_hump_spike} for some examples), interpreted to be unresolved Rossby modes (the \textit{hump}) mechanically excited by deviated flows caused by stellar spots (the spike) at intermediate to high latitudes \citep{2018MNRAS.474.2774S}. Rossby modes were first reported in intermediate-mass stars by \citet{2016A&A...593A.120V}, as they studied the core rotation of a sample of $\gamma$ Dor stars. They have been reported in other works such as: \citealt{2019MNRAS.487..782L, 2020MNRAS.491.3586L}.

\subsection{Alternative explanation for rotational modulation }

Recently, \citet{LeeSaio2020} and \citet{Lee2021} proposed a different explanation for rotational modulation of light curves of early-type stars, which does not involve the presence of stellar spots generated by magnetic fields. 
Overstable convective (OsC) modes in rapidly rotating early-type stars could resonantly excite low-frequency g~modes in the envelope. If the stellar core rotates slightly faster than the envelope and the amplitudes of the g~modes are significant at the photosphere, they would cause brightness variability and therefore be observed as rotational modulations (i.e. the spike).

In addition, \citet{Lee2021} suggested that OsC modes resonating with g~modes in the envelope can play an essential role in carrying angular momentum from the core to the envelope. A small amount of radial differential rotation is a critical ingredient in the \citet{LeeSaio2020} and \citet{Lee2021} hypothesis, as OsC modes would not excite g~modes in uniformly rotating stars \citep{LeeSaio2020}. Previous studies in which core and envelope rotation rates were measured through photometric data found that up to about 10 per cent differential rotation between convective core and envelope is not rare in $\gamma$ Doradus stars \citep{2021MNRAS.502.5856S}. 

In the context of our \textit{hump \& spike} sample of stars, we investigate the possibility that the spike could in some cases correspond to g~modes confined near the core and resonantly excited by the convective core motions. In this case, the light curves of our \textit{hump \& spike} stars would be modulated by the convective core rotation. 

In this article, we address both scenarios and discuss the corresponding implications. More precisely, we investigate whether the spike feature could be either: (1) caused by rotational modulation due to stellar spots caused by magnetic fields or (2) evidence that OsC modes couple with g~modes in the envelope, causing the light curve to be modulated with the convective core rotation. 

Under the assumption that spots induced by magnetic fields cause the spike modulation, the first goal of the present work is to investigate whether the stars in our sample could sustain a magnetic field. We will test two scenarios: failed-fossil field \citep{2013MNRAS.428.2789B} and dynamo-generated field (\citetalias{Cantiello_2019}). 

The second goal of this study is to test whether the spike is evidence of convective core rotation, using predictions from \citet{Lee2021}. 
If the frequency of the spike is proven to correspond to the frequency of a g~mode that is resonantly excited by OsC modes, then our results would help better understand the convective core rotation in intermediate-mass stars as well as the internal mixing. We describe observations, data reduction and analysis in Section\,\ref{sec:obs} and discuss our results in Section\,\ref{sec:results}.The implications of our results under both the magnetic field and the OsC modes hypotheses are presented in Section \ref{sec:discussions}. Our conclusions and future work are presented in Section\,\ref{sec:conclusion}.

\section{Observations and data analysis}
\label{sec:obs}

In this work, we use the Kepler long cadence data \citep{2010ApJ...713L..79K}, which are excellent for our study given the long time coverage, and the fact that the signals of interest are found at low frequencies ($<20\,{\rm d}^{-1}$). 
The light curves from each quarter were downloaded from KASOC (Kepler Asteroseismic Science Operations Center)\footnote[1]{\url{kasoc.phys.au.dk}}. For our analysis, we used the PDC (Pre-search Data Conditioning) data, as they were free of specific systematic errors, removed by the Kepler Science
Operations Center Pipeline \citep{2010SPIE.7740E..1UT}. We visually inspected each PDC light curve and compared it to the corresponding SAP light curve (Simple Aperture Photometry). We concluded that the \textit{hump \& spike} feature remained unaltered after the time series were processed by the PDC module of the Kepler data analysis pipeline. The data from KASOC is barycentrically corrected, and the light curve files have the Barycentric Reduced Julian Date (BRJD, BJD-2,400,000.0) as time units \citep{2016ksci.rept....2V} \footnote[2]{\url{archive.stsci.edu/kepler/manuals/Data\_Characteristics\_Handbook\_20110201.pdf}}.

The flux was converted from units of photo-electrons per second ($e^{-}{\rm s}^{-1}$) to ppm, after which the quarterly data were stitched into a single time series for each star. We computed a Fourier spectrum for each stitched time series using the \textit{lightkurve} package \citep{2018ascl.soft12013L}. 

\subsection{Sample selection}
\label{sec:sample_select}

\begin{figure}
	\includegraphics[width=\columnwidth]{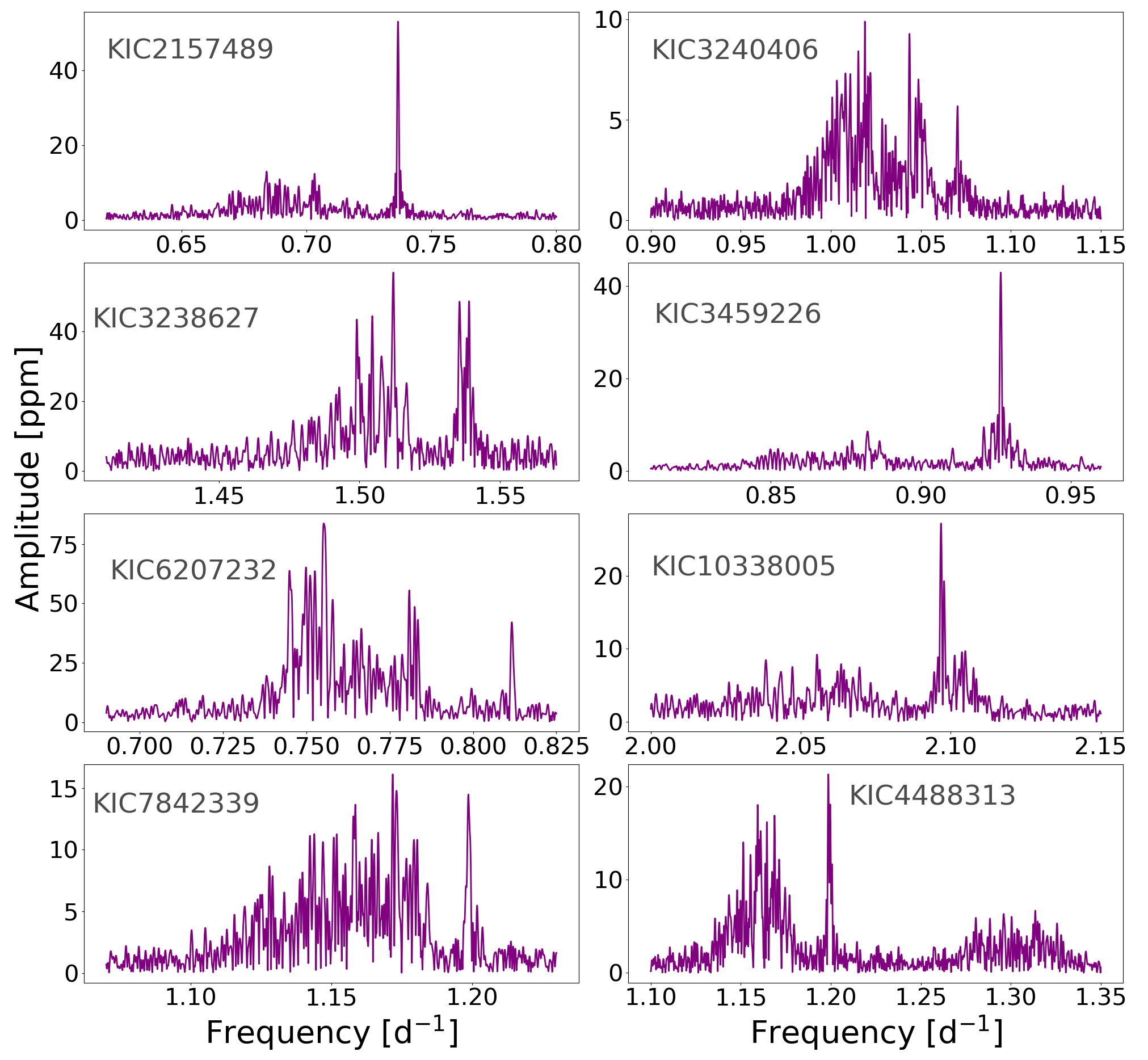}
    \caption{ A selection of  \textit{hump \& spike} features, illustrating their common aspects and amplitude and frequency diversity. }
    \label{fig:zoo_hump_spike}
\end{figure}

The target sample initially comprised 94 stars that were visually inspected and found to exhibit the \textit{hump \& spike} feature or were reported in \cite{2013MNRAS.431.2240B, 2014MNRAS.441.3543B, 2017MNRAS.467.1830B,2015MNRAS.448.1378B}. The temperature range on which we concentrated to find targets was $6500 < T_{\rm eff} < 10 000$\,K. An additional set of 115 stars ($7500 < T_{\rm eff} < 10 000$\,K) from \citet{2020MNRAS.492.3143T} was included in our study.  Another four stars were added to our sample from \citealt{2019FrASS...6...46M} and by visually inspecting the data products available from \citet{2021ApJS..255...17S}.
A total of 213 stars that exhibited the \textit{hump \& spike} feature were analysed. However, in this article, we concentrate on 162 stars that did not show signatures of binarity (either from photometry or literature). This selection was performed in order to study a homogeneous sample of stars. We note that some stars in our final sample exhibit a hump after the spike (e.g., KIC\,4488313 in Fig. \ref{fig:zoo_hump_spike}). This feature is similar to one identified by \citealt{2018MNRAS.474.2774S}, who suggested
that it is probably the result of prograde g modes (see figure 8 in \citealt{2018MNRAS.474.2774S}). This feature will be addressed in a future work that will concentrate more on Rossby modes.

\begin{figure}
	\includegraphics[width=\columnwidth, height=135pt]{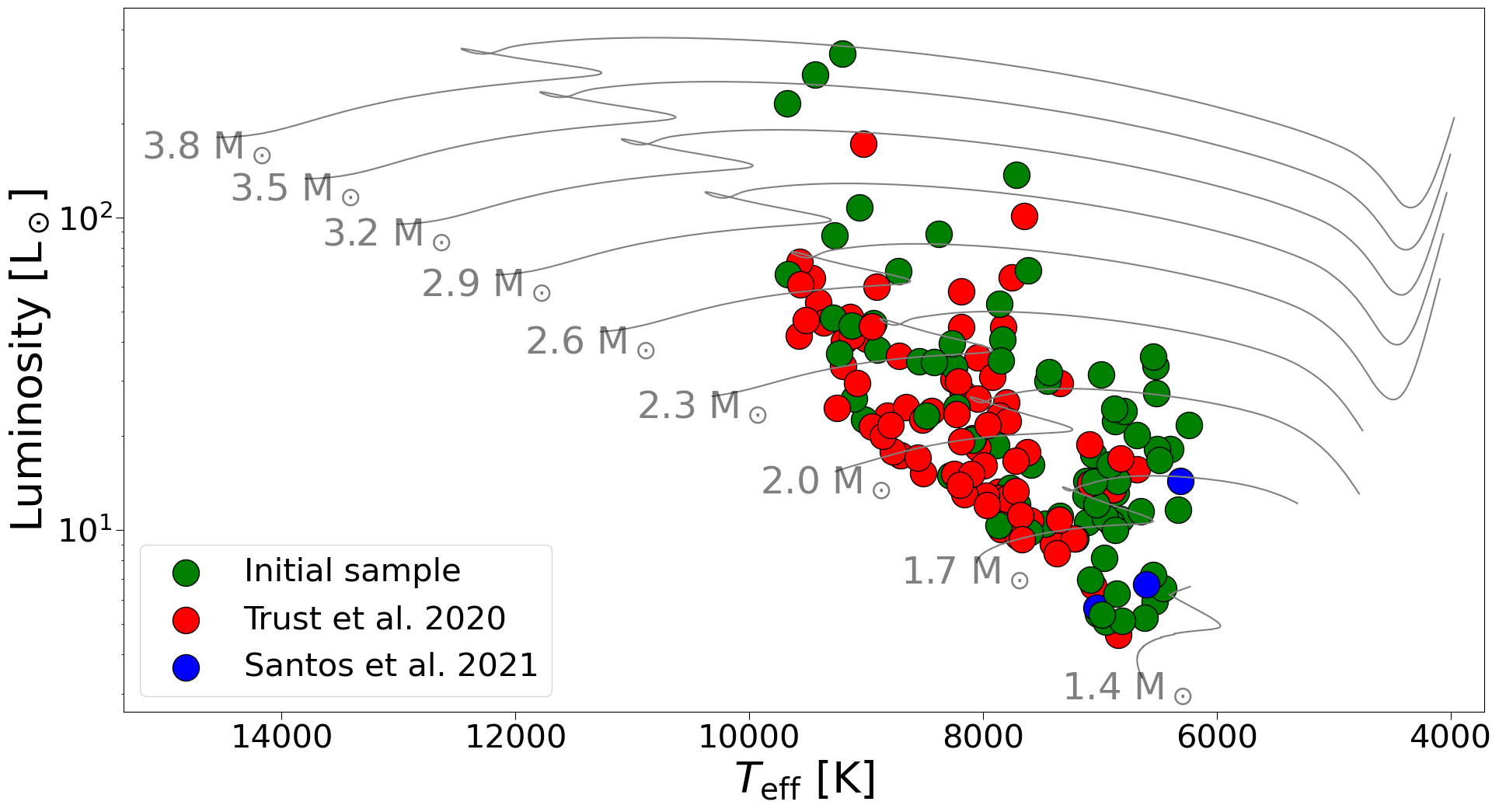}
    \caption{Stars from our sample placed in an HR diagram. The source for luminosity and $T_{\rm eff}$ values is described in section \ref{sec:lum_teff}. The colour code represents the source from which the target originates, as described by the legend. More information in section \ref{sec:sample_select}. Warszaw-New\,Jersey evolutionary tracks ($Z=0.012$, \citealt{2004A&A...417..751A}) are displayed in the background for guidance only.}
    \label{fig:clean_HR}
\end{figure}

In Fig.\,\ref{fig:clean_HR}, we show an HR~diagram illustrating our final sample. Stellar evolutionary models for the mass range $1.4 - 3.8 {\rm M}_{\odot}$, computed with Warszaw-New\,Jersey models \citep{1998ASPC..135..232V, 1999AcA....49..119P, 1998A&A...333..141P} are shown in the background for guidance only, computed with solar metallicity \citep{2004A&A...417..751A} and with no rotation. The full sample of stars can be seen in  Fig.\,\ref{fig:HRD_all} in Appendix \ref{sec:apx1}, with the excluded stars marked in orange squares and purple triangles while the targets used in the current work are marked in green circles.

The feature of interest is, as mentioned, the \textit{hump \& spike} with a few examples shown in Fig.\,\ref{fig:zoo_hump_spike}, and a more detailed view in the left panel of Fig.\,\ref{fig:ex_gauss_fit_spike}. The following subsection describes how the parameters of the spikes were extracted.

\subsection{Spike parameter extraction}
\label{sec:spike_param}

For most stars in our sample, the PDC light curves did not need additional de-trending or corrections. Only $\sim$10 per cent of PDC light curves still contained some non-astrophysical signals, to which some minor corrections were applied. A clear example of a light curve that needed such corrections can be seen in Fig.\,\ref{fig:corr_lk_h_spike}, where only the jump in flux at around BRJD 56\,000) was corrected by fitting three low-order polynomials to the data. The corrections were done keeping in mind that over-fitting might introduce additional and unnecessary signals in the data, so the order of the polynomials was kept as low as possible. In this example, three second-degree polynomials sufficed. As seen in the lower panel of Fig.\,\ref{fig:corr_lk_h_spike}, the Fourier spectra of both polynomial-corrected data and PDC data are over-plotted, respectively, the astrophysical signal (\textit{hump \& spike} feature) is unaltered after the correction, but the noise at lower frequencies is reduced. The applied corrections eliminated only the non-astrophysical signals, evidently showing that the peaks at low frequencies are indeed associated with the jump in flux that occurs in the time series at BRJD $\sim$56\,000 in the upper panel of Fig.\,\ref{fig:corr_lk_h_spike}. We computed also a Fourier spectrum without the data that contain the jump in flux (in the time range 55904 and 56015), coloured in grey in the lower panel of Fig.\,\ref{fig:corr_lk_h_spike}. The small differences in the respective Fourier spectra indicate the non-astrophysical origin of the jump. We conclude that the correction applied did not introduce additional non-astrophysical signals.  Moreover, these minor corrections helped obtain a higher signal-to-noise ratio (SNR) for the studied spikes and/or humps. 

For each star in our sample, a simple 1D Gaussian model was fitted to the spike (and its harmonics) using \textit{astropy} \citep{astropy:2013, astropy:2018}. Spike identification was done semi-automatically. We divided the Fourier spectrum into equal sections and identified the peaks with \textit{scipy} \citep{2020SciPy-NMeth}. Only peaks with a SNR $\geq 4$ were accepted as significant. After identifying a significant peak, we defined a region in the Fourier spectrum (blue shaded region in Fig.\,\ref{fig:spike_selection}, spike selection of KIC\,8385850) and fitted a simple Gaussian. Manual spike identification was necessary due to the variety in the \textit{hump \& spike} profiles. As seen in Fig.\,\ref{fig:zoo_hump_spike}, the topology of the \textit{hump \& spike} feature changes from star to star. In some cases, the spike was not well separated from the hump or was a broad feature rather than a sharp narrow peak.

The noise level for the SNR calculations was estimated to be the median of the amplitude values between a range selected at frequencies higher than $\sim$15-20$\,\mathrm{d^{-1}}$, where no astrophysical signal could be identified. 
The frequency range for SNR calculations was selected for each star with the same tool as depicted in Fig.\,\ref{fig:spike_selection}. Usually the SNR is determined around the extracted peak. However given the proximity of the hump (both at lower frequencies and in some case at higher frequencies), it was not possible to determine the noise level around the spike. We therefore chose to use the noise level at higher frequencies which we expect to mirror the white noise in the data set.

\begin{figure}

	\includegraphics[width=\columnwidth]{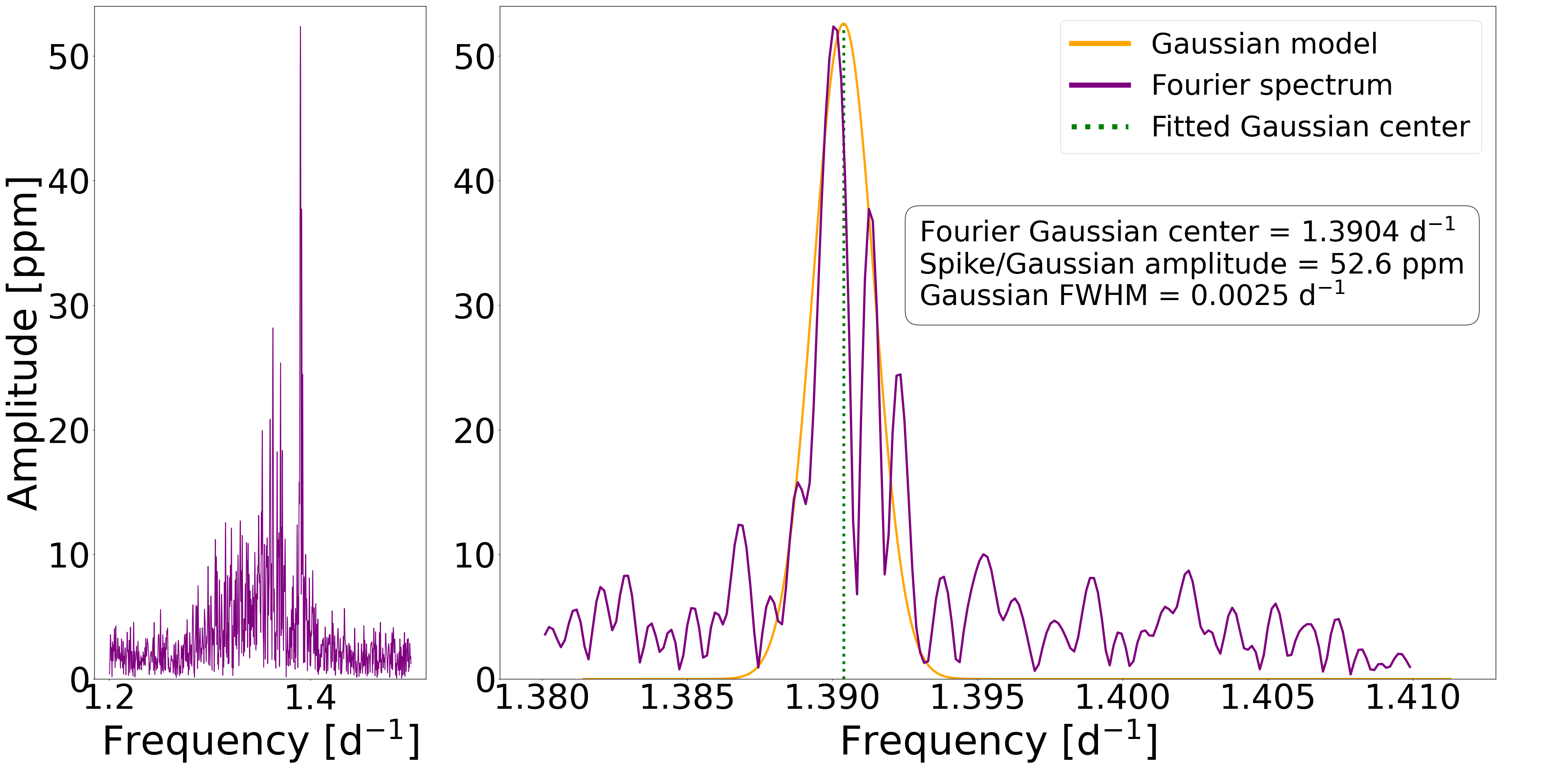}
    \caption{ Example of a Gaussian fit on a spike feature, KIC\,4921184. }
    \label{fig:ex_gauss_fit_spike}
\end{figure}

\begin{figure}

	\includegraphics[width=\columnwidth]{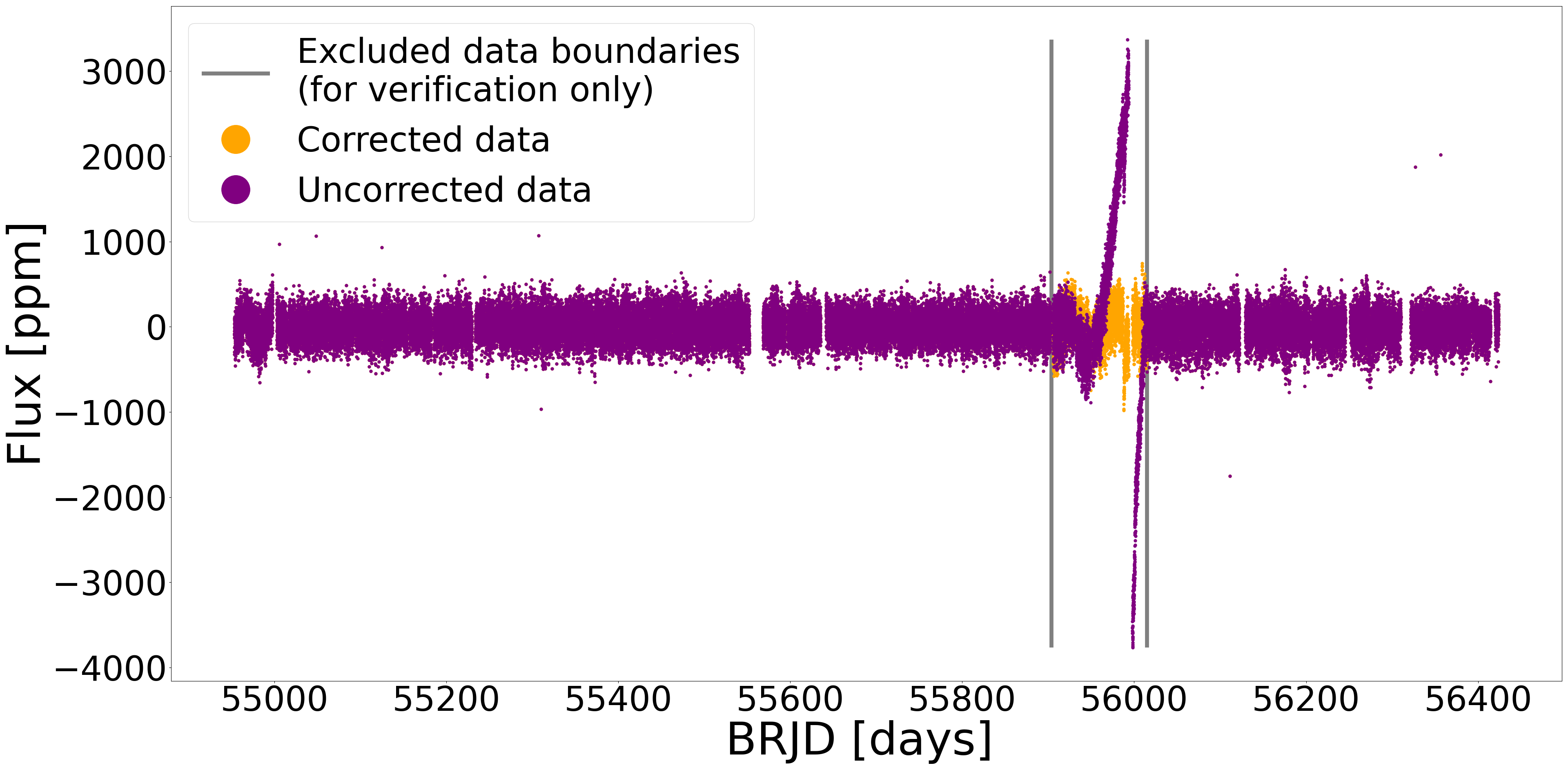}	
	\hfill
	\includegraphics[width=\columnwidth]{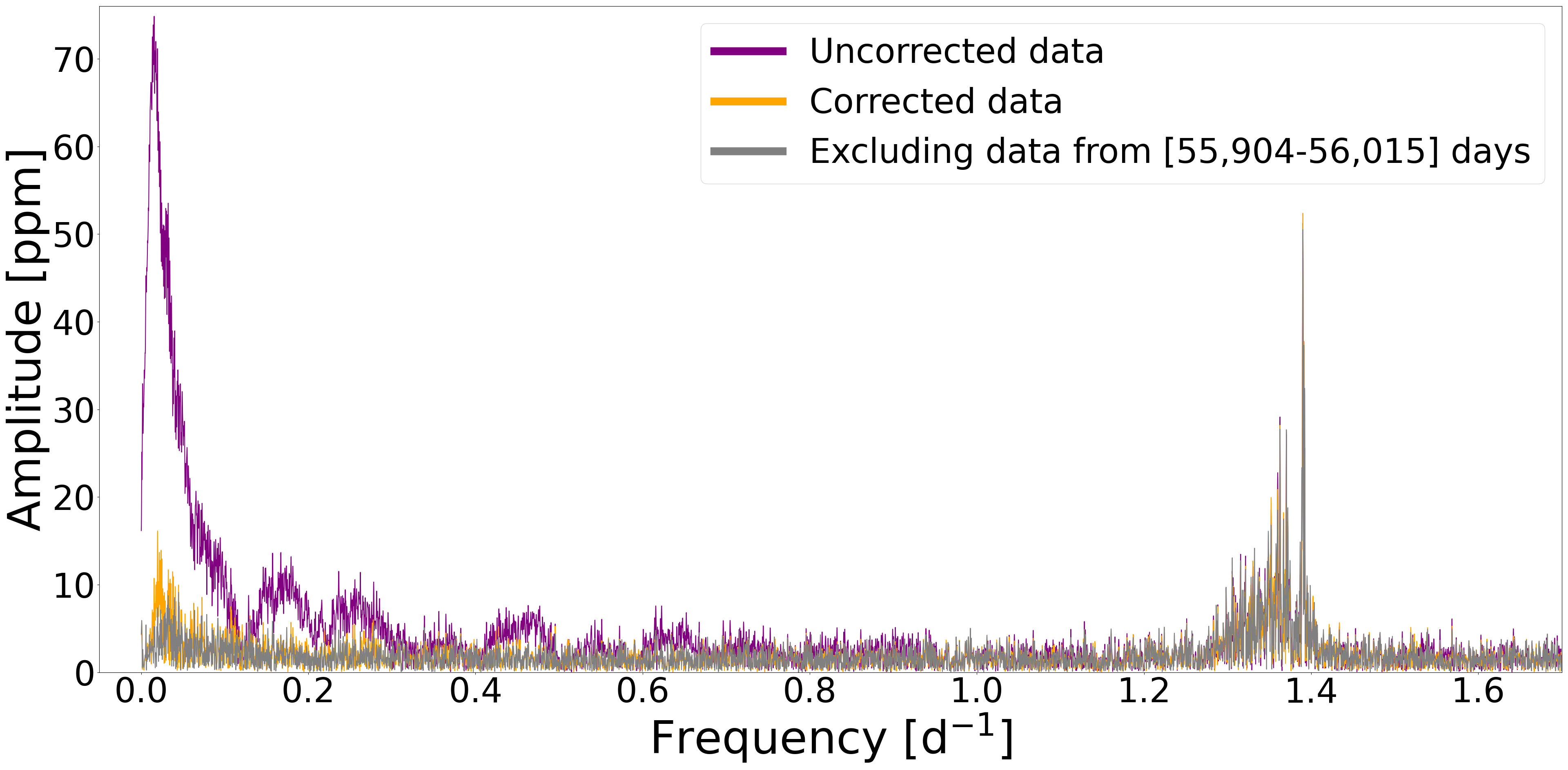}
    \caption{\textit{Upper panel:} PDC Light curves of KIC\,4921184: before the correction was applied (purple), after non-astrophysical signal was removed (orange). See Section\,\ref{sec:spike_param} for more details. \textit{Lower panel:} Fourier spectra of corrected (orange) and uncorrected (purple) light curves. The grey Fourier spectrum was computed with the full data set excluding the region inside the two grey vertical lines from upper panel. There are small differences between the Fourier spectrum of the corrected data and the one computed without the data between BRJD 55904 and 56015 days.}
    \label{fig:corr_lk_h_spike}
\end{figure}

\begin{figure}
	\includegraphics[width=\columnwidth]{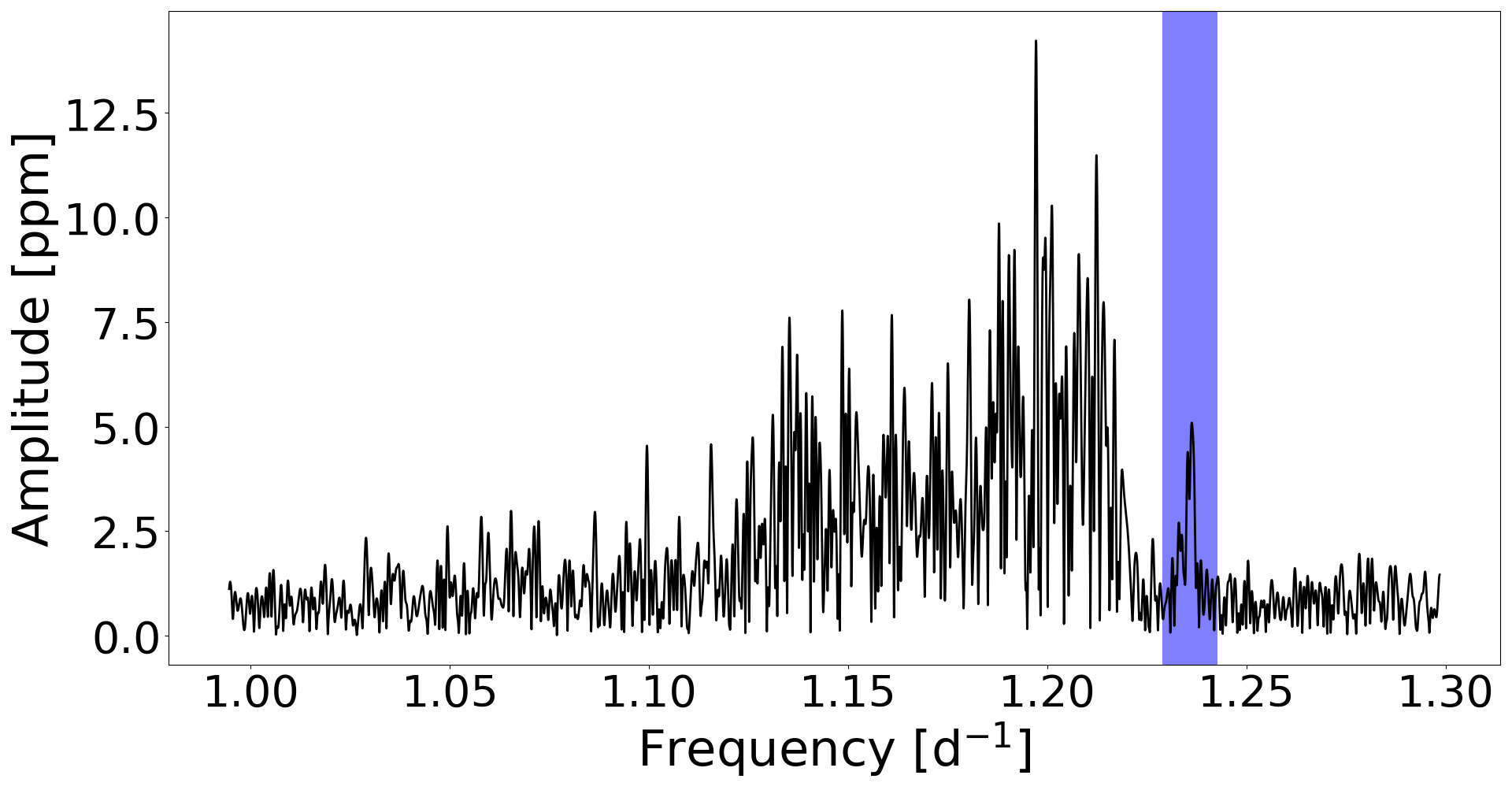}
    \caption{Illustration of how a frequency region was selected for the Gaussian fitting - KIC\,1873552. }
    \label{fig:spike_selection}
\end{figure}

Similar to \cite{2020MNRAS.492.3143T}, we define the value for the spike amplitude to be the highest amplitude value in the selected spike region. This is a robust way to determine the photometric variability in the 4-year Kepler data.
The uncertainties in spike amplitudes were calculated with equation \ref{eq:sigma_ampl}, which was adapted from \citealt{1995A&A...293...87K}.

\begin{equation}\centering
    \begin{aligned}
    \mathrm{\sigma_A} = \frac{\sqrt{\pi}\,SNR}{2}
    \end{aligned}
     \label{eq:sigma_ampl}
\end{equation}

where SNR is the noise level in the Fourier spectrum.

We note that the spike amplitudes could be underestimated. If the spike signal is due to stellar spots, the amplitude would depend on the angle at which the spot is observed, which in turn depends on the stellar inclination and the latitude of the spot. Determining the intrinsic value of the spike amplitude requires knowledge of the stellar inclination, which is out of the scope of the current work.

Almost all stars (121) have four years of \textit{Kepler} data. However a few stars have missing quarters, as indicated in Table \ref{tab:spike_param1}. In the latter case, gaps in the data will introduce alias peaks. This will increase the uncertainty of $f_{rot}$, as it would not be possible to discern between the correct and alias peaks. Furthermore if the length of the data set is shorter, the precision of $f_{rot}$ will again decrease. For example, KIC\,5456027 is the only star that has 12 missing quarters, as seen in Table \ref{tab:spike_param1} ($f_{rot} = 1.242$ $\rm d^{-1}$, $\sigma f_{rot} = 0.0039$ $\rm d^{-1}$). The uncertainty on $f_{rot}$ is on average, an order of magnitude larger than for other stars, meaning that the uncertainty in determining the rotation frequency increases with the existence of gaps in the data.

\begin{figure}

	\includegraphics[width=\columnwidth, height=135pt]{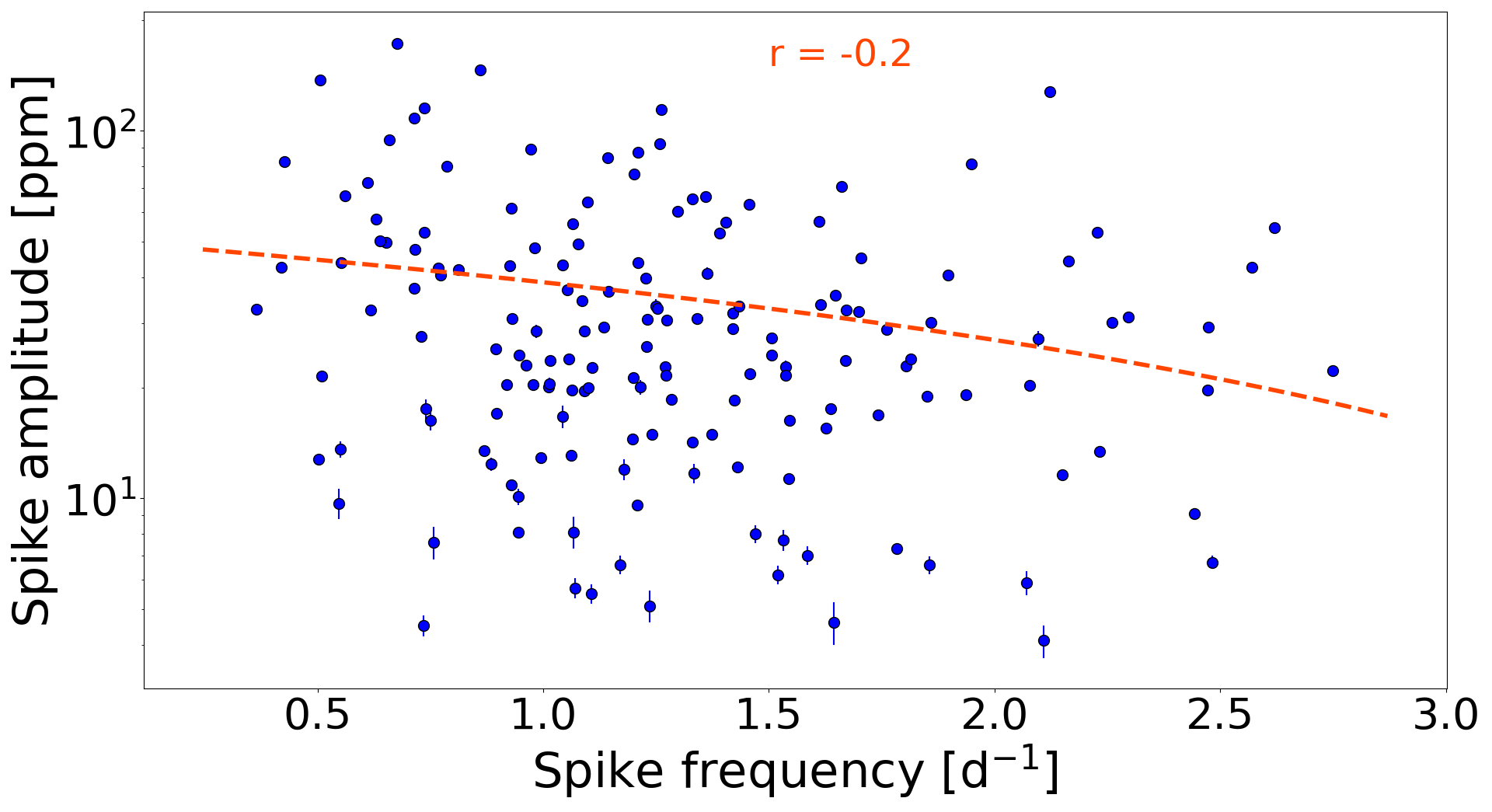}
    \caption{Spike amplitude as a function of spike frequency; if not visible the uncertainties in both quantities are smaller than the symbols.}
    \label{fig:spk_ampl_spk_freq}
\end{figure}

\noindent The spike frequency and amplitude and the associated uncertainties for each star can be found in Table \ref{tab:spike_param1}. Figure \ref{fig:spk_ampl_spk_freq} shows a weak anti-correlation between the frequency and amplitude of the spike, suggesting that rapidly rotating stars tend to have slightly lower spikes.

\begin{figure}
	\includegraphics[width=\columnwidth]{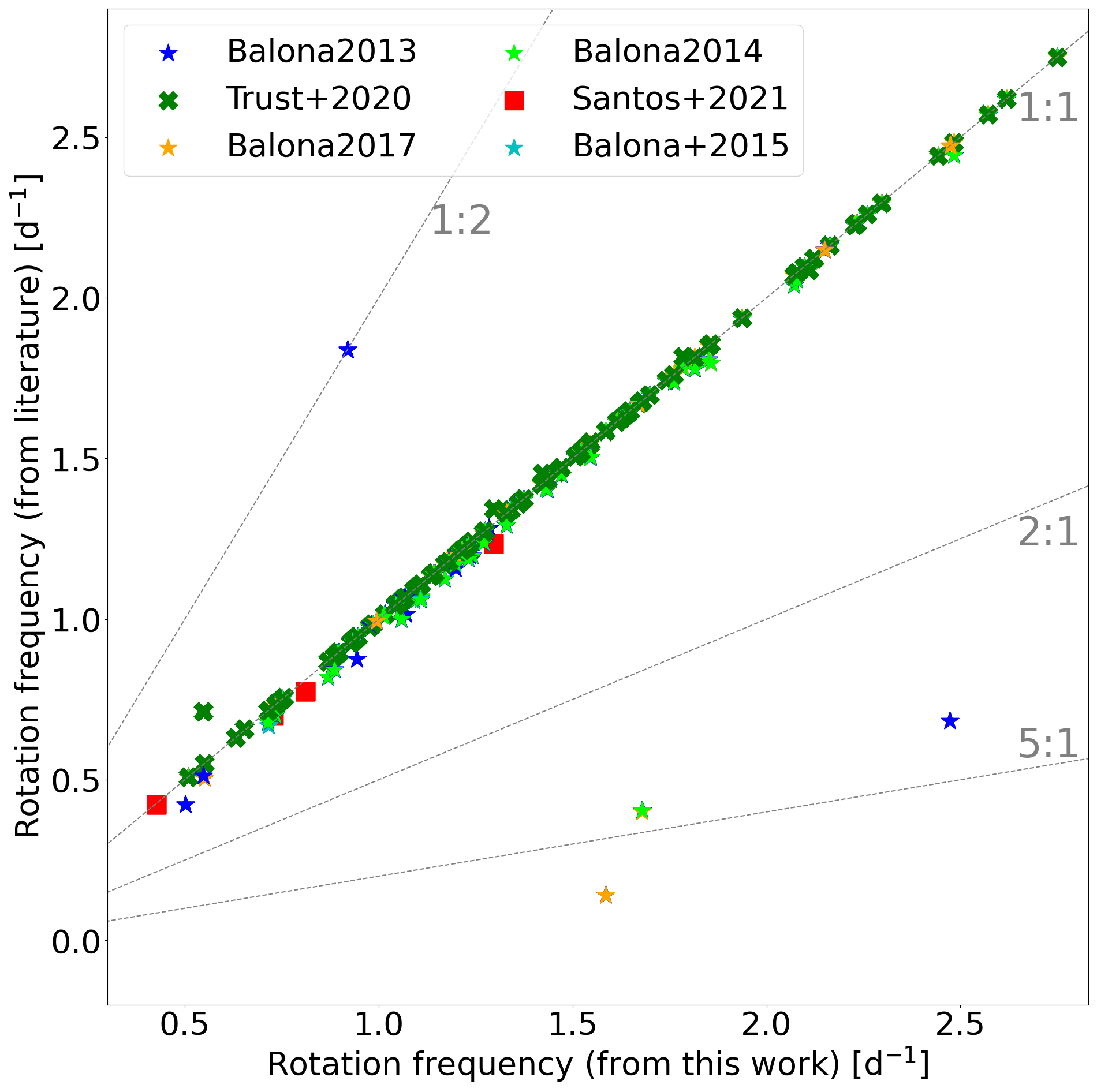}
    \caption{ Comparison between rotation frequency from this work versus literature values: \citet{2013MNRAS.431.2240B,2014MNRAS.441.3543B, 2015MNRAS.448.1378B, 2017MNRAS.467.1830B, 2020MNRAS.492.3143T, 2021ApJS..255...17S}. }
    \label{fig:frot_comparison}
\end{figure}

\begin{figure}
	\includegraphics[width=\columnwidth]{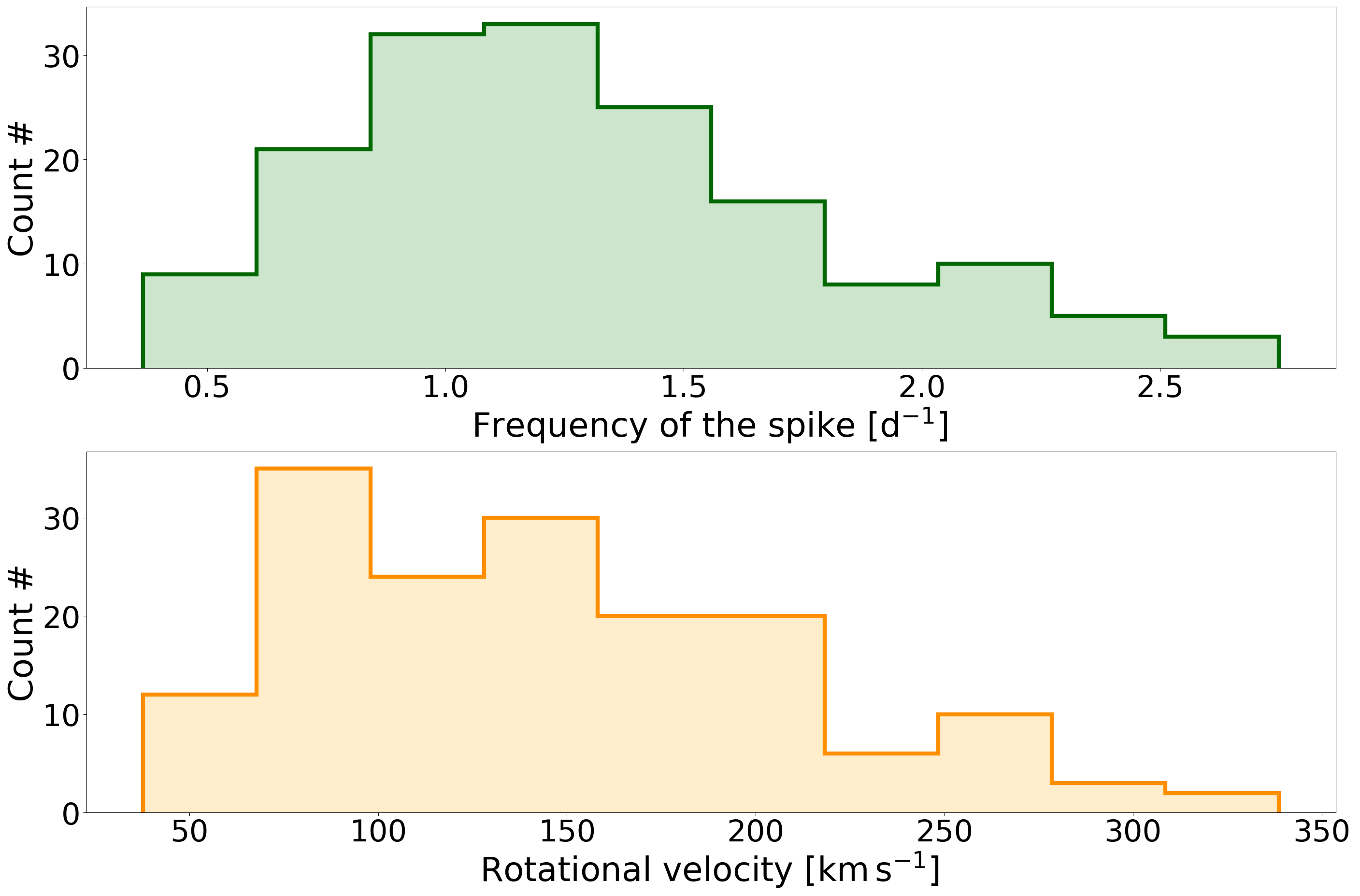}
    \caption{\textit{Upper panel:} Histogram of the rotation frequencies extracted for our sample of stars (spike frequency). \textit{Lower panel:} Histogram of rotational velocities, calculated with equation \ref{eq:vrot}.}
    \label{fig:hist_frot_vrot}
\end{figure}

The right panel of Fig.\,\ref{fig:ex_gauss_fit_spike} shows a fitting example for KIC\,4921184. Values for the rotation frequency were compared to those in literature, as seen in Fig.\,\ref{fig:frot_comparison}. For details regarding the few differences, see Section \ref{sec:results}. The upper panel of Fig.\,\ref{fig:hist_frot_vrot} depicts a distribution of the rotation frequency values, which lie between 0.28 and 2.75~$\rm d^{-1}$, corresponding to rotation periods between 3.6 and 0.4~d. This highlights the short rotation periods of our stars.

\begin{figure}
	\includegraphics[width=\columnwidth, height=135pt]{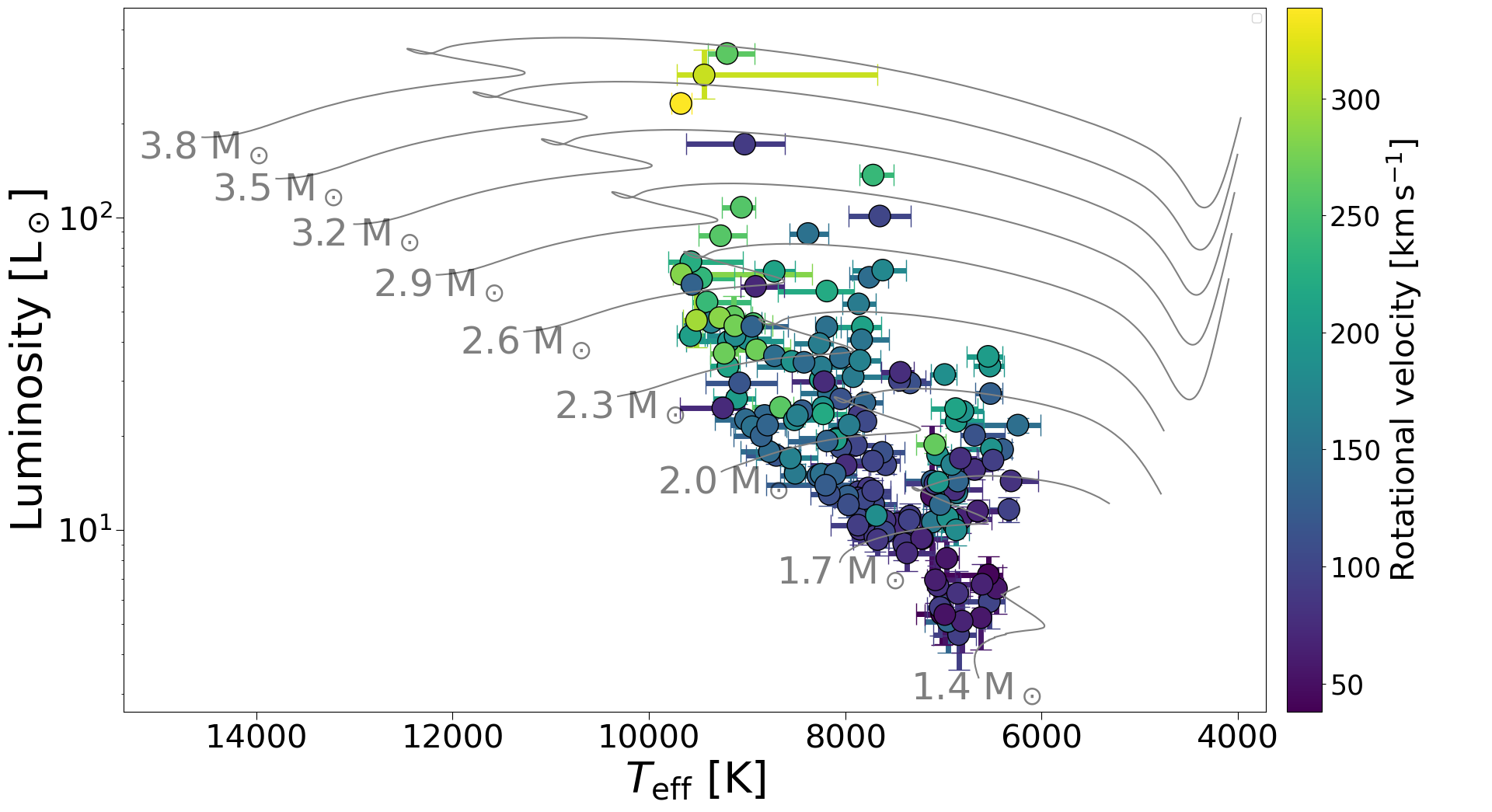}
    \caption{HR diagram with the symbol colour correlated to the rotational velocity, calculated using the spike frequency and equation \ref{eq:vrot}. The source for luminosity and $T_{\rm eff}$ values is described in section \ref{sec:lum_teff}. Warszaw-New\,Jersey evolutionary tracks ($Z=0.012$, \citealt{2004A&A...417..751A}) are displayed in the background for guidance only.}
    \label{fig:HRD_vrot}
\end{figure}

In Fig.\,\ref{fig:corr_plots}, panel (d1) depicts the relation between the rotational velocity and the spike amplitude for each target in our sample. The rotational velocity was calculated as:

\begin{equation}
        V_{\rm rot}\,(\mathrm{km \ s^{-1}}) = \frac{2 \ \pi \ R_s  \  F_{\rm rot}}{86400}
\label{eq:vrot}
\end{equation}

\noindent where $R_s$ is the stellar radius in km and $F_{\rm rot}$ is the rotational frequency in $\mathrm{d^{-1}}$, corresponding to the frequency of the main spike. For 93 targets from our sample, the radius values and the associated uncertainties were extracted from \textit{Gaia} DR2 \citep{2018A&A...616A...1G}. For the remaining 69 stars, the radius values were taken from \citet{2020AJ....159..280B}. The uncertainty in stellar radius has the highest contribution to the uncertainty in the rotational velocity. The values obtained for the rotational velocities are depicted in a histogram in the lower panel of Fig.\,\ref{fig:hist_frot_vrot} and are also listed in Table \ref{tab:spike_param1} where the radius values and the associated uncertainties can also be found. \citet{2018Ap&SS.363..260C} reported rotation periods of 513 stars, out of which 394 are within the same $T_{\rm eff}$ range as our targets. Their reported rotational frequencies are similar to our sample, suggesting that the \textit{hump \& spike} stars do not rotate differently from other A and F stars. Fig. \ref{fig:HRD_vrot} depicts the rotational velocities with respect to their location in the HR diagram, confirming  that more massive, hotter stars rotate faster. We briefly discuss the importance of rapid rotation of some of our stars in section \ref{sec:harmonic_rotation}, as it is an important ingredient in the OsC theory.

\subsection{Luminosity and $T_{\rm eff}$ values}
\label{sec:lum_teff}

Luminosity values for 156 targets from our sample were extracted from \citet{2019MNRAS.485.2380M} (supplementary data). For the remaining six stars not found in \citet{2019MNRAS.485.2380M}, we assigned luminosity values from \citet{2020AJ....159..280B} (four stars) and from \text{Gaia} DR2 (two stars). Last column in Table \ref{tab:spike_param1}, denotes the source of the stellar parameters as explained in the caption. 

For obtaining the luminosity values, \citet{2019MNRAS.485.2380M} used the \text{Gaia} DR2 parallaxes and uncertainties, $g$ apparent magnitudes (derived from $g_{\rm KIC}$ and $r_{\rm KIC}$ magnitudes and calibrated to the SDSS scale), extinctions and their uncertainties from the \textit{dustmap} python package (Bayestar 17 reddening map) and bolometric corrections with \textit{isoclassify} python package. For a more in-depth explanation of how the values were obtained, see \citet{2019MNRAS.485.2380M} and references therein. 

The luminosity values for the four stars, extracted from the \citet{2020AJ....159..280B} catalogue, were derived from isochrones and broadband photometry, \textit{Gaia} DR2 parallaxes, and spectroscopic metallicities. See \citet{2020AJ....159..280B} for a more in-depth understanding of how these values were obtained. The effective temperatures ($T_{\rm eff}$) for all 162 stars were extracted from \textit{Gaia} DR2. 

\subsection{Harmonics of the main spike}
\label{sec:harm_spike}

\begin{figure}

	\includegraphics[width=\columnwidth]{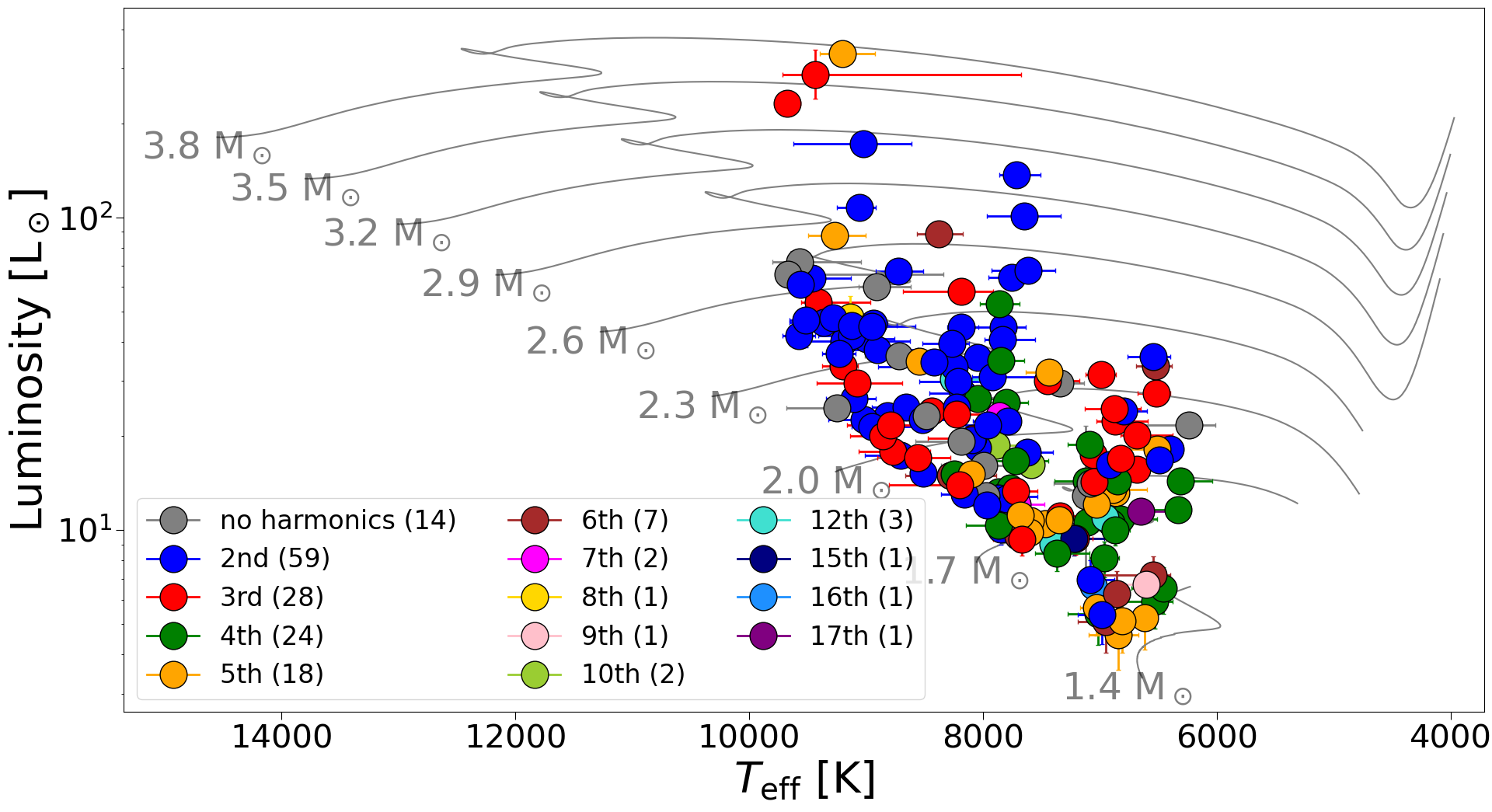}
    \caption{ HR diagram with the symbol colour indicating the highest harmonic detected in the Fourier spectrum. The number in the paranthesis next to each item from the legend indicates the number of stars in the sample that have the respective highest harmonic. The source for luminosity and $T_{\rm eff}$ values is described in section \ref{sec:lum_teff}. Warszaw-New\,Jersey evolutionary tracks ($Z=0.012$, \citealt{2004A&A...417..751A}) are displayed in the background for guidance only.}
    \label{fig:hrd_harm}
\end{figure}

The majority of the stars in our sample exhibit one to several harmonics of the spike frequency, which can be interpreted as a sign of a non-sinusoidal signal. Only 14 stars do not have any detected harmonics of the main spike. The colour code in Fig.\,\ref{fig:hrd_harm} is dictated by the highest harmonic found in the Fourier spectrum. There seems to be no apparent correlation between the presence or lack of harmonics and the stellar parameters. In the legend of Fig.\,\ref{fig:hrd_harm}, the number of stars that possess specific harmonics can be found. We focus more on the significance and importance of harmonics for our stars in Section \ref{sec:harmonic_rotation}. 

\subsection{Time-series analysis}
\label{sec:time_series}

\begin{figure}
	\includegraphics[width=\columnwidth]{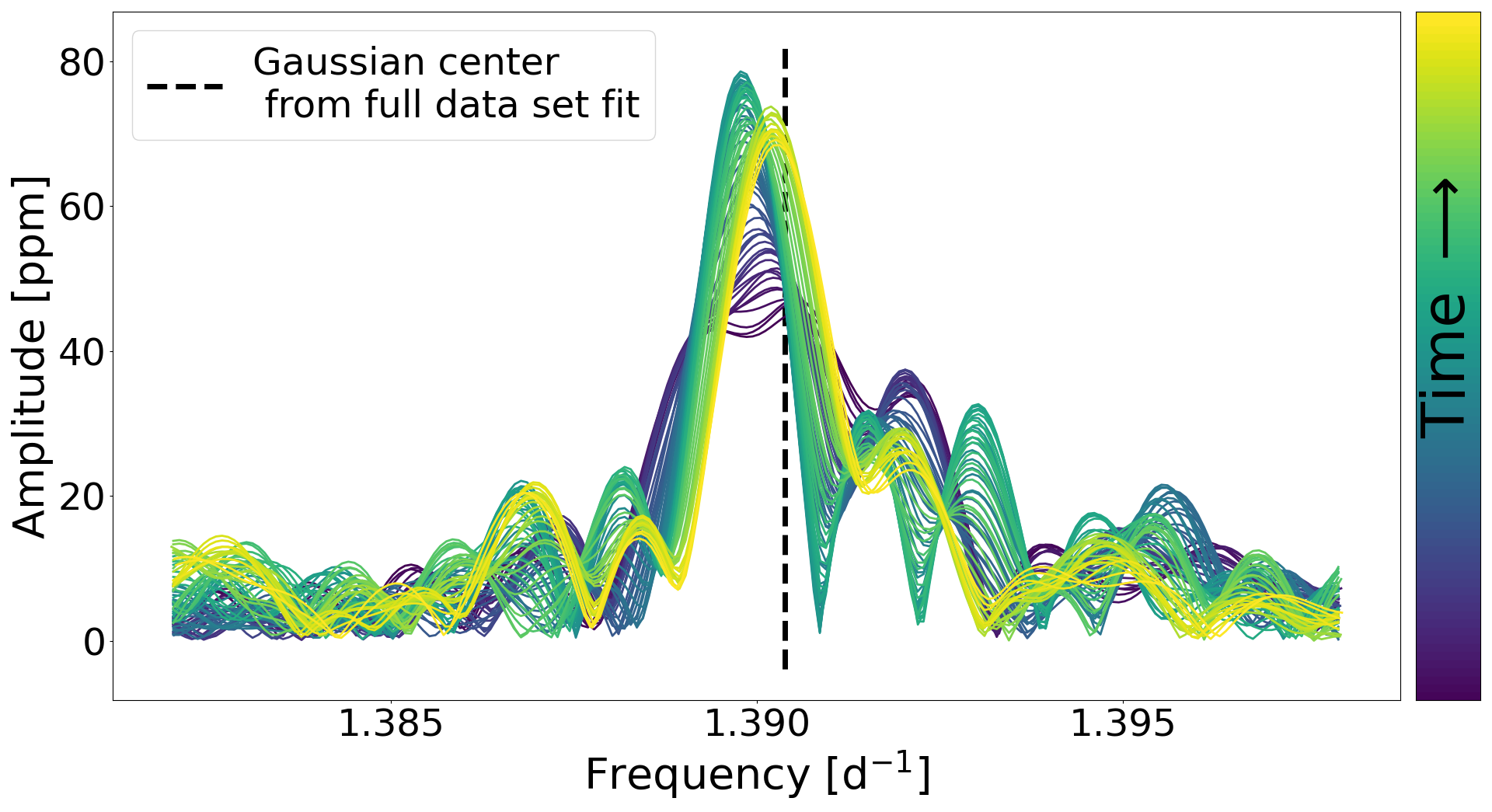}
    \caption{KIC\,4921184 Fourier spectra computed with subsets of the full Kepler data colour coded with respect to time. The yellow Fourier spectrum is computed with the most recent Kepler data.  Black vertical line at $\sim 1.39 \mathrm{d^{-1}}$ denotes the frequency of the spike, obtained by fitting a Gaussian to the full 4-year Kepler data. The amplitude of the spike in the full data set is $52 \pm 0.5 $ ppm.}
    \label{fig:time_series2}
\end{figure}

In order to verify the temporal behaviour of the periodic signals, we performed a time series analysis on the data, focusing on temporal changes of the spike, as shown in Fig.\,\ref{fig:time_series2}, where Fourier spectra of subsets of the KIC\,4921184 Kepler data are displayed. The average duration of each sub-data sets is 800\,d and the average overlap 791\,d.  Each Fourier spectrum was computed with the \textit{lightkurve} package \citep{2018ascl.soft12013L}. The colour coding in Fig.\,\ref{fig:time_series2} represents the time when the photometric data from the data subsets were acquired, the yellow Fourier spectrum being computed with the most recent Kepler data. The Gaussian centre obtained from fitting the spike in the Fourier spectrum of the full data set (see also Fig. \ref{fig:ex_gauss_fit_spike}) is depicted in Fig.\,\ref{fig:time_series2} as a vertical black dotted line at the corresponding frequency. Fig.\,\ref{fig:time_series2} clearly shows the amplitude and structure of the spike change in time.
The values for computing  Fig.\,\ref{fig:time_series2} (800\,d sub-data sets, overlapping 791\,d) were chosen as a trade-off between resolution, SNR and having enough subsets to illustrate the changing character of the spike. However, we quantify this temporal change using other methods as described in section \ref{sec:time_decay}.

The Fourier spectrum computed with the full Kepler data set of KIC\,4921184 is shown in Fig.\,\ref{fig:kic4921184_fourier}. The harmonics of the spike (f1 = 1.39 $\mathrm{d^{-1}}$) are also visible (2f1, 3f1, 4f1, 5f1 = 2.78, 4.17, 5.56, 6.95 $\mathrm{d^{-1}}$). The presence of harmonics is indicative of non-sinusoidal brightness variations, typical for stellar spots. Furthermore, the changing shape of the main spike, as seen in Fig.\,\ref{fig:time_series2}, and the double-wave light variation in the middle panel of Fig.\,\ref{fig:bpf_timeseries} suggest the presence of variable stellar spots (e.g., \citealt{2018Ap&SS.363..260C}).  
In general, a double-wave variation indicates the presence of spots that are in anti-phase, causing a stronger $2^{\rm nd}$ harmonic \citep{2017A&A...599A...1S}. However, for KIC\,4921184, the amplitude of the $2^{\rm nd}$ harmonic is just as high or slightly lower than that of the main spike in some of the Fourier spectra computed with subsets of the full Kepler data, suggesting the presence of non-permanent spots.

\subsection{Spike lifetime}
\label{sec:time_decay}
\begin{figure}
	\includegraphics[width=\columnwidth, height=135pt]{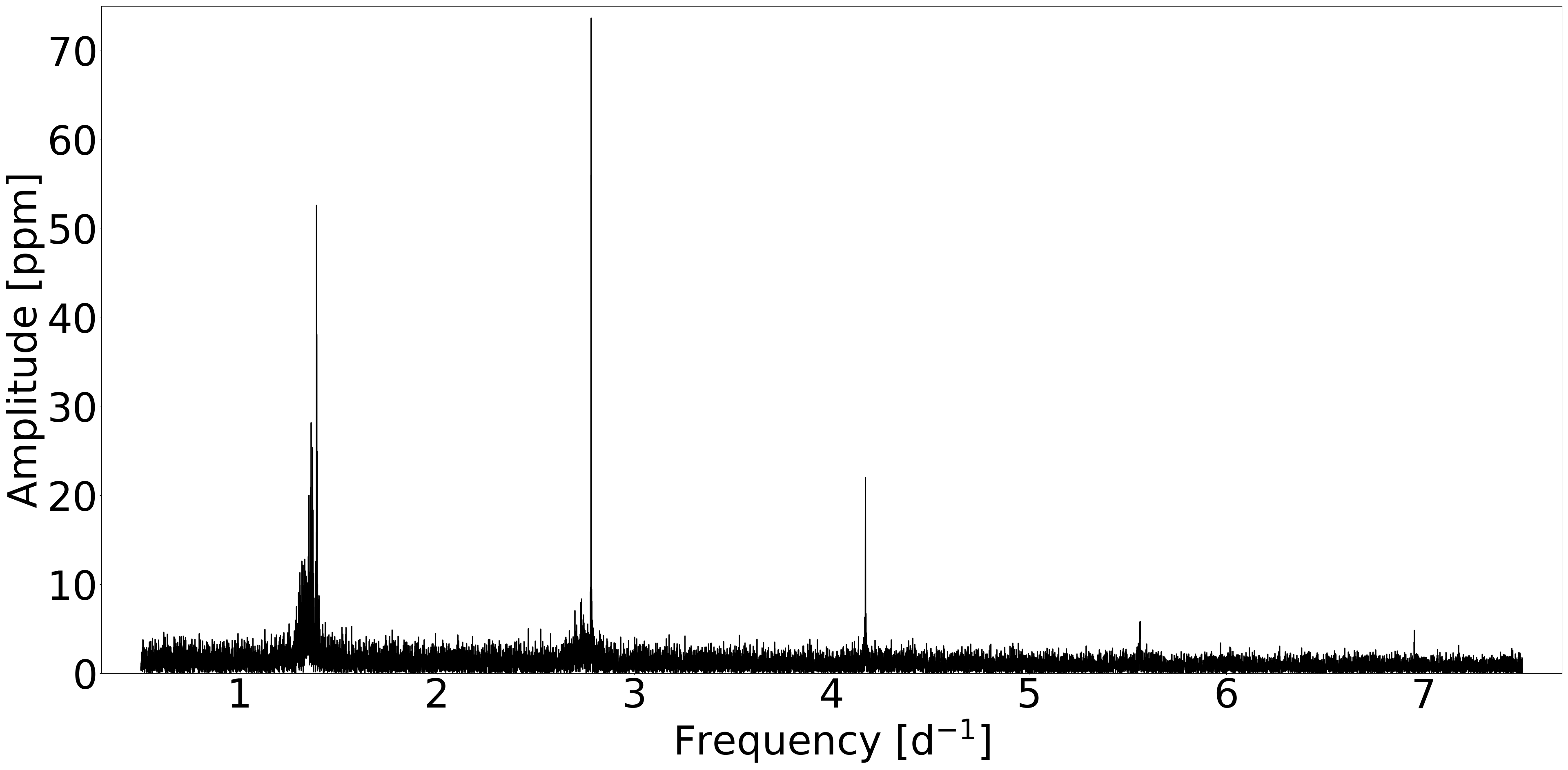}
    \caption{ KIC\,4921184 Fourier spectrum with the  main spike (f1 = 1.39 $\mathrm{d^{-1}}$) and its harmonics (2f1, 3f1, 4f1, 5f1 = 2.78, 4.17, 5.56, 6.95 $\mathrm{d^{-1}}$).}
    \label{fig:kic4921184_fourier}
\end{figure}

Given the changing nature of the spike (Fig.\,\ref{fig:time_series2}), we searched for a way to quantify its evolution in time. It has been suggested that the decay time-scales of the autocorrelation functions (ACFs) of light curves are related to the lifetimes of stellar spots \citep{2017MNRAS.472.1618G,2020MNRAS.492.3143T,2021MNRAS.508..267S}. Given that one of the goals of our work was to probe whether the spike is evidence for magnetic fields, we decided to compute the ACFs based on which we could extract the spike lifetime. We only concentrated on the signal induced by the spike and its harmonics, requiring a band-pass filtered time series. We calculated the sum of all  Fourier components in the frequency bands around the spike and the harmonic frequencies, $n\,f_{\rm rot} \pm \Delta f$,  where $\Delta f$ is the frequency range around $n$ times the spike frequency. $\Delta f$ depended on the broadness of the spike/harmonic, which varied from case to case; $n$ takes values between 1 and the order of the highest detected harmonic. For example, for KIC\,4921184 (Fig. \ref{fig:kic4921184_fourier}) $n=5$. The Fourier components were determined using an iterative sine-wave fitting process, which was stopped once all the signal within the $n\,f_{\rm rot} \pm \Delta f$ frequency range was recovered.

\begin{figure}
		\includegraphics[width=\columnwidth, height=135pt]{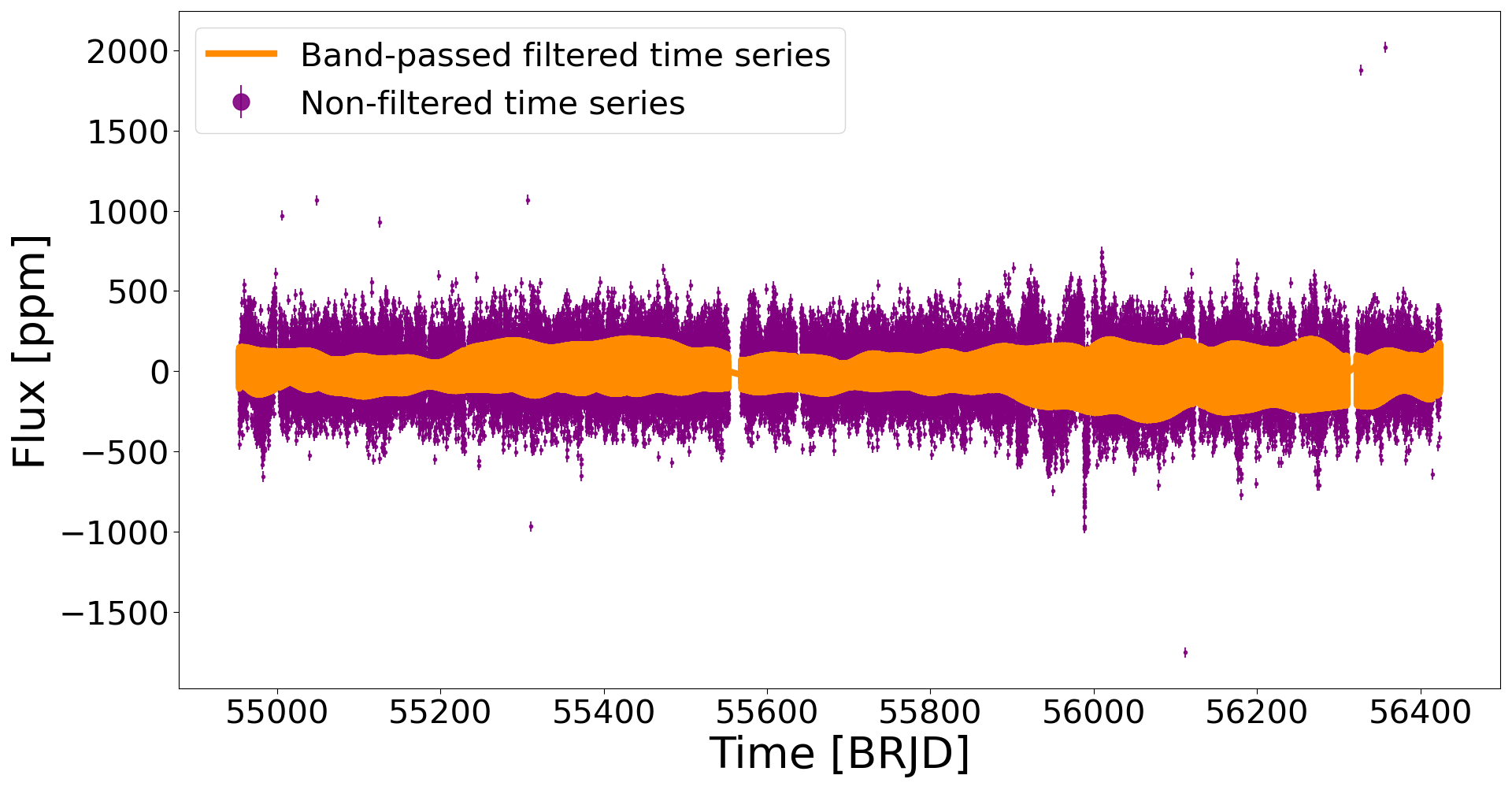}
	\hfill
    		\includegraphics[width=\columnwidth, height=135pt]{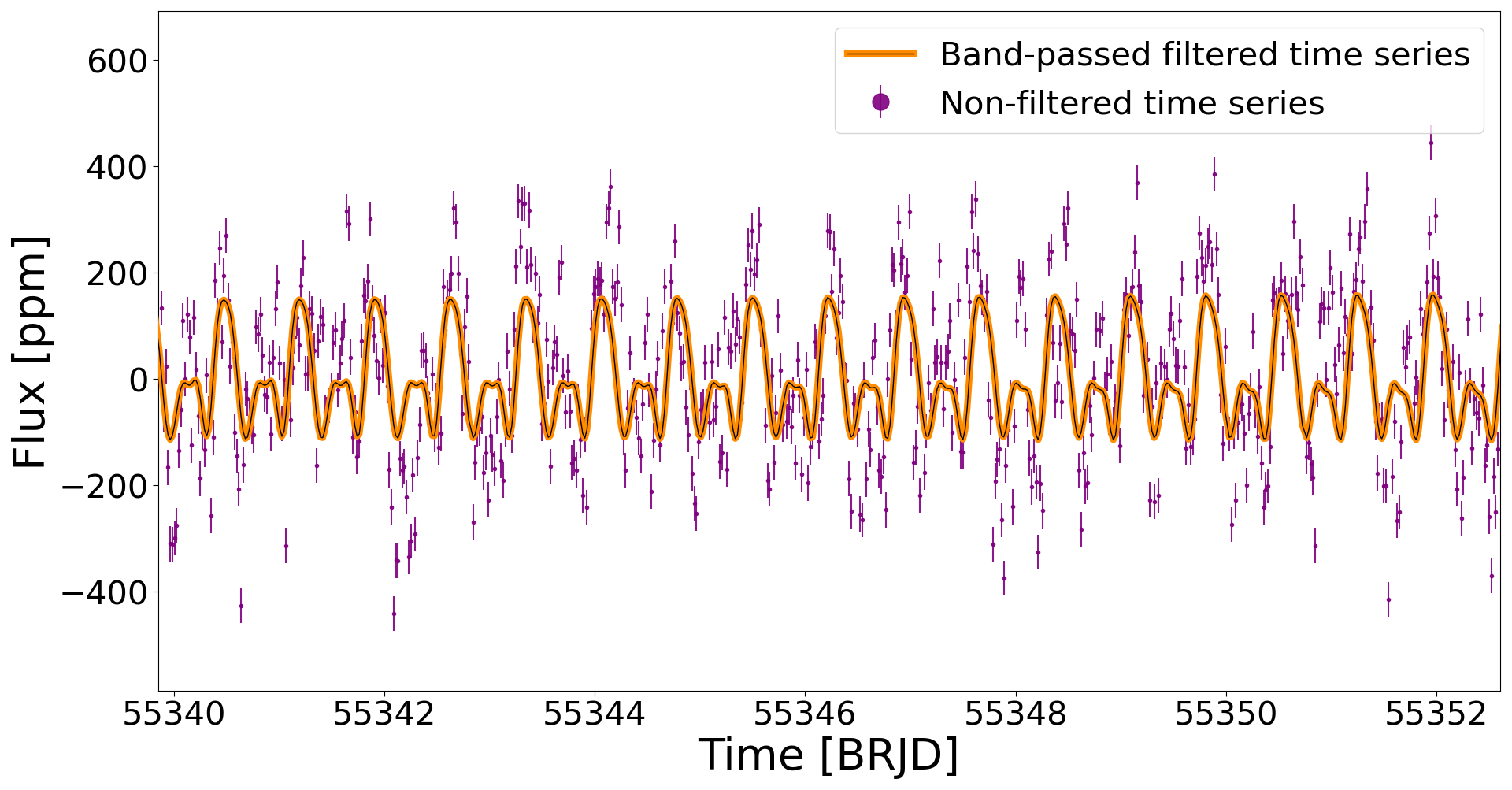}
	\hfill
    		\includegraphics[width=\columnwidth, height=135pt]{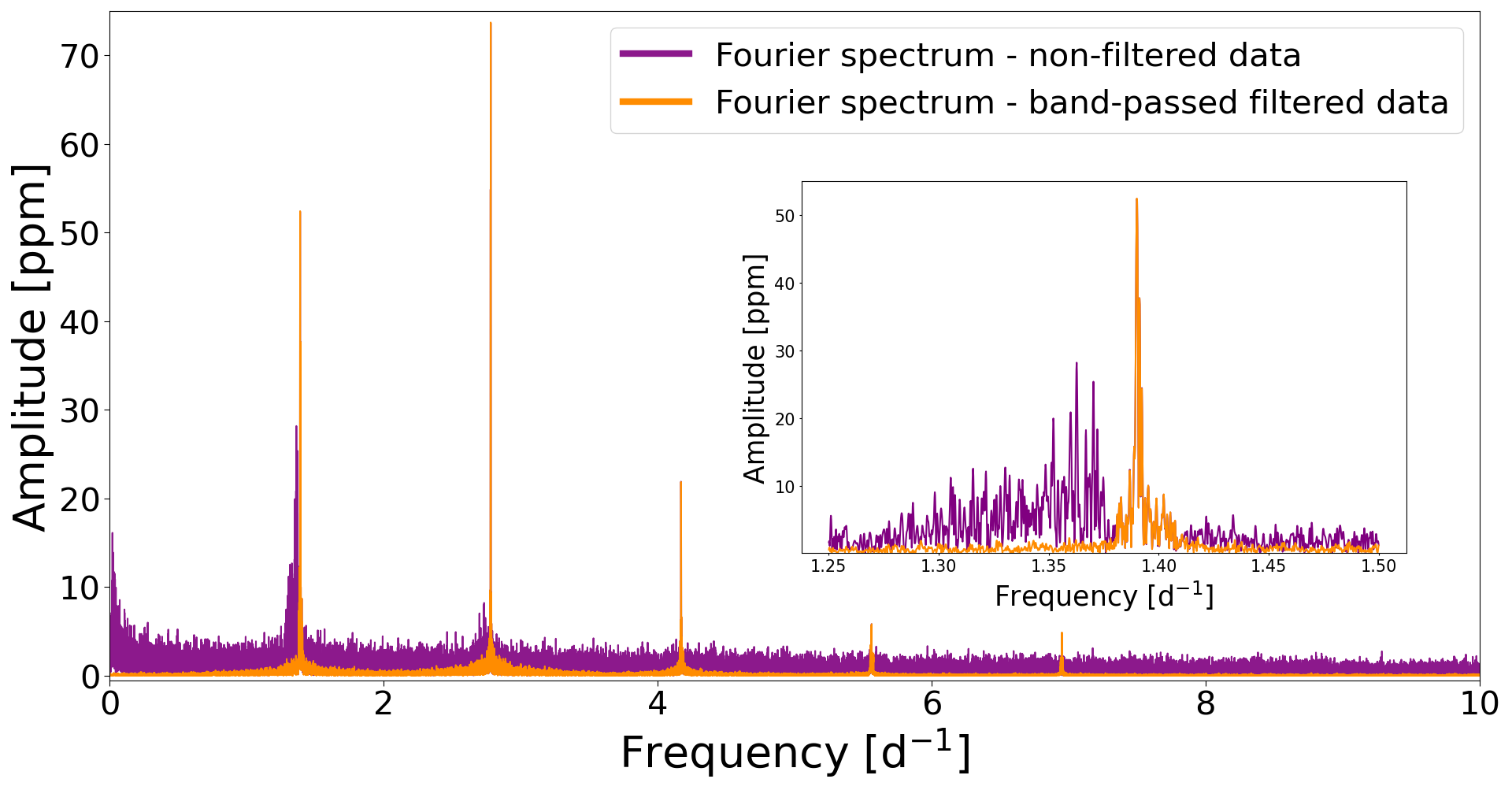}
\caption{\textit{Upper panel:} Band-pass filtered time series and the original Kepler data for KIC\,4921184. \textit{Middle panel:} Zoom in of Figure a), the shape of the non-sinusoidal signal induced by the spike and its harmonics are depicted in orange, while the original data are the purple dots. \textit{Lower panel:} Fourier spectrum of both the original data set (purple) and the band passed filtered time series (orange), where only the spike and harmonic signal is contained. The insert shows a zoom in on the first spike.}
\label{fig:bpf_timeseries}
\end{figure}

Fig.\,\ref{fig:bpf_timeseries} shows an example of the band-pass filtered time series. In the upper panel of Fig.\,\ref{fig:bpf_timeseries} the contribution of the spike and its harmonics to the brightness variability is illustrated over the complete Kepler data, while the mid-panel shows a zoom-in on the non-sinusoidal signal induced by the spike and its harmonics. The lower panel of Fig.\,\ref{fig:bpf_timeseries} shows the difference in Fourier spectra computed with the full and band-passed filtered time series. The insert in the lower panel depicts a zoom in on the first spike and its hump. 

\begin{figure}
	\includegraphics[width=\columnwidth, height=135pt]{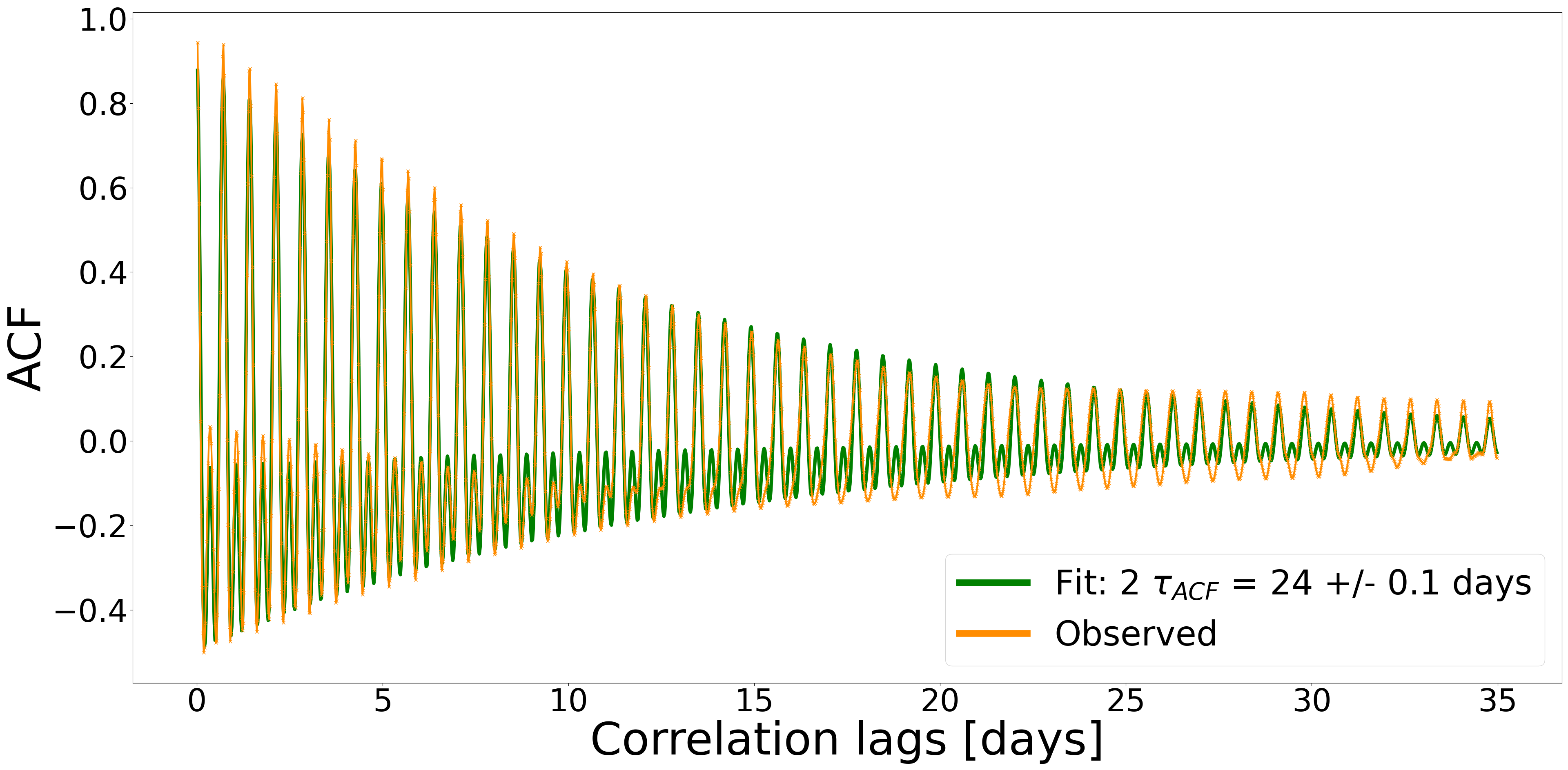}
    \caption{Autocorrelation function of the band-passed time series KIC\,4921184. The orange line is the ACF of band-passed filtered time series. The green line is a fit obtained with equation \ref{eq:time_decay}.}
    \label{fig:ACF_example}
\end{figure}

After computing the ACF of the band-passed filtered time series (orange curve in Fig.\,\ref{fig:ACF_example}, obtained with \textit{numpy} package \citealt{harris2020array}), the result was fitted with an exponential model (green line), described by equation \ref{eq:time_decay}, using least-squares minimization \citep{2020SciPy-NMeth}.

\begin{equation}\centering
    \begin{aligned}
        y(t) = \left(e^{\frac{-t}{\tau_{\rm ACF}}}\right) \left[y_0 + A cos  \left( \frac{2\pi t}{P_{A\!C\!F}}\right) + B cos \left(\frac{4\pi \!t}{P_{A\!C\!F}}\right) \right]
    \end{aligned}
     \label{eq:time_decay}
\end{equation}

\begin{figure}
	\includegraphics[width=\columnwidth]{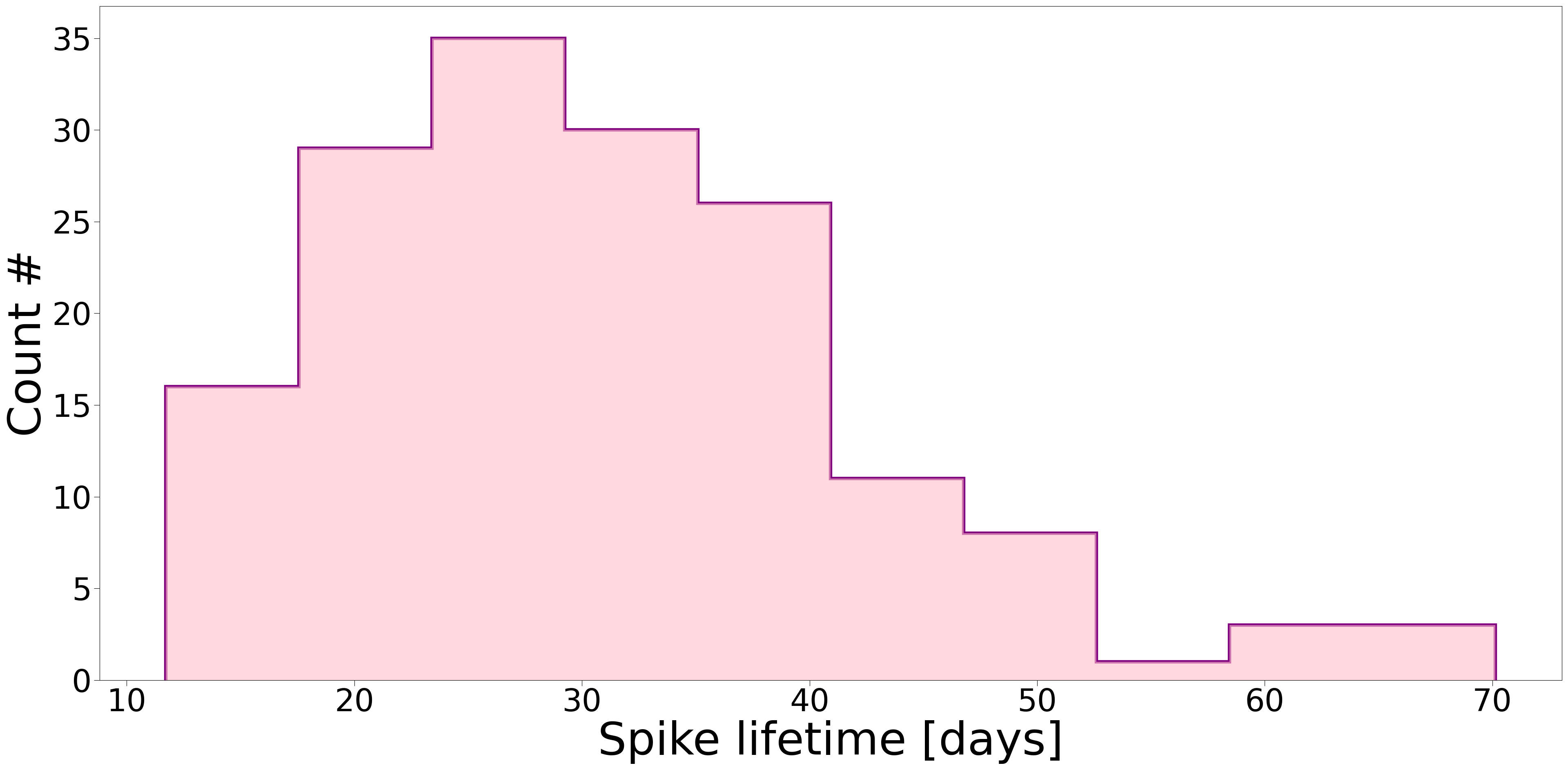}
    \caption{Spike lifetime distribution ($2 \,\tau_{\rm ACF}$) obtained with equation \ref{eq:time_decay}.}
    \label{fig:hist_t_decay}
\end{figure}

\noindent In equation \ref{eq:time_decay}, introduced by \citep{2017MNRAS.472.1618G} to approximate the starspot lifetime of Kepler stars, \textit{t} represents the correlation time lags in days obtained as $t = N\, \Delta T $, with $\Delta T$ being the median time difference between consecutive observations in the light curve and \textit{N} is the number of observed data points in a light curve. $\tau_{\rm ACF}$ is the decay time-scale of the magnetic features (stellar spots) and $P_{\rm ACF}$ is the rotation period of the star. $A, B$ and $y_0$ are not associated with physical stellar properties, but constants used in the fitting algorithm of above mentioned equation. All parameters in equation \ref{eq:time_decay} were left free in the fitting process. Fig.\,\ref{fig:hist_t_decay} depicts the spike lifetimes distribution, obtained using equation \ref{eq:time_decay}. The values obtained are similar to what \citet{2020MNRAS.492.3143T} reported for a sample of stars with common targets.

The accuracy of the method was tested on synthetic time series with known spike lifetimes. We simulated 400 time series with spike lifetimes of 15, 30, 60 and 120\,d (100 time series for each value). The total length of each time series is 1500 days, with a 30 minutes cadence. The synthetic time series did not contain noise. The period of the signal was one day for all time series. Amplitude variability (excitation and damping) was calculated in a similar way as simulated for solar-like p-mode oscillations (see \citealt{2006MNRAS.365..595D}). The average uncertainty was $\sim$27 per cent for time series with a spike lifetime of 15, 30, 60\,d, and $\sim$42 per cent for 120\,d, respectively. In terms of accuracy, the method yielded values $\sim$$1-2$ per cent away from the injected ones for 15, 30, and 60\,d, and $\sim$$21$ per cent for 120\,d, respectively. 

\begin{figure}
	\includegraphics[width=\columnwidth]{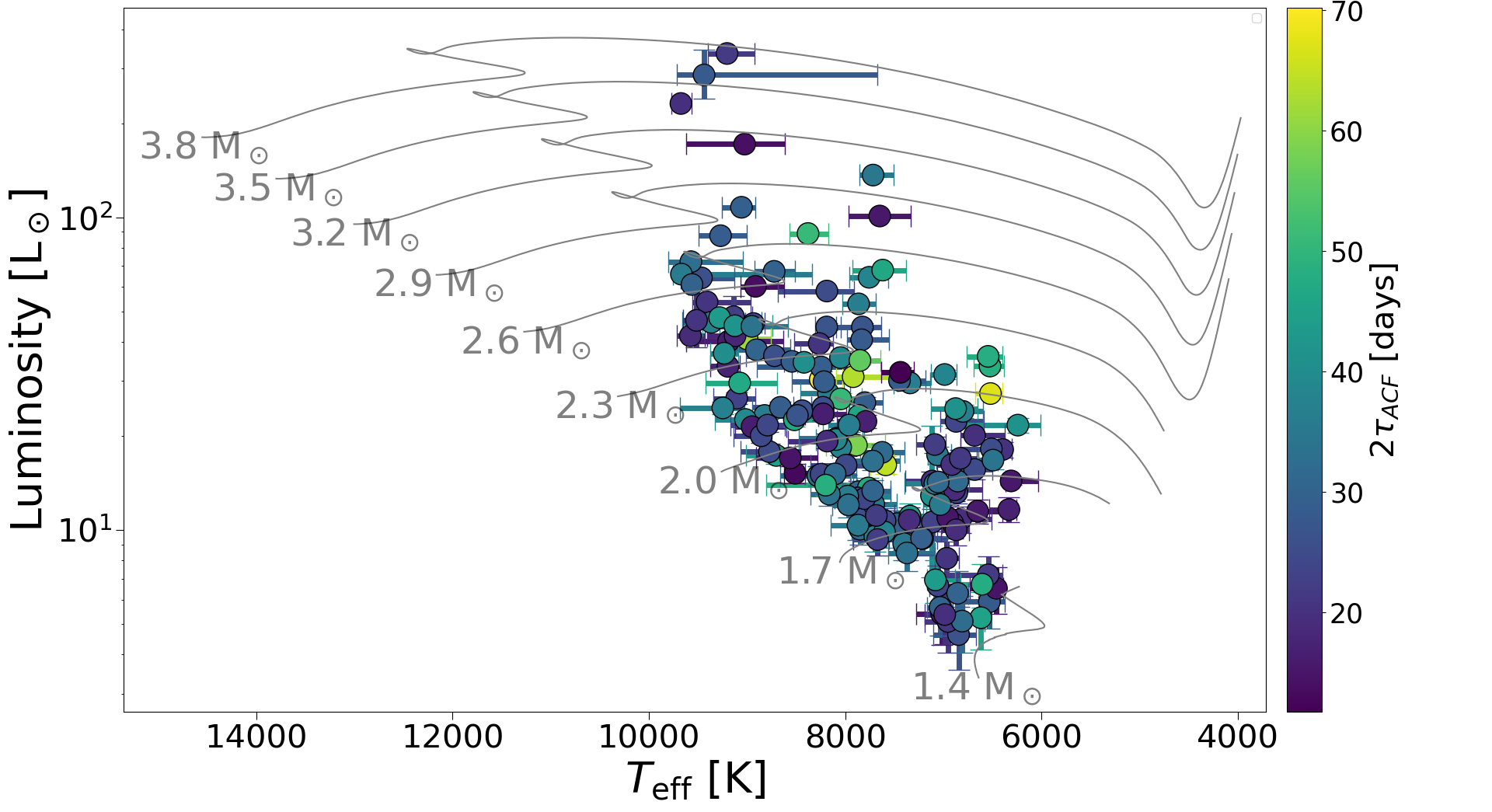}
    \caption{HR diagram with symbols correlated with the spike lifetime as calculated in section \ref{sec:time_decay}. The source for luminosity and $T_{\rm eff}$ values is described in section \ref{sec:lum_teff}. Warszaw-New\,Jersey evolutionary tracks ($Z=0.012$, \citealt{2004A&A...417..751A}) are displayed in the background for guidance only.}
    \label{fig:HRD_time_decay}
\end{figure}

\citealt{2021MNRAS.508..267S} suggested that the exponential model underestimates decay time-scales in stellar spot simulations. They introduced an alternative to the model proposed by \citealt{2017MNRAS.472.1618G}. We find, however, that  the ACFs in this work are better fitted with the \citealt{2017MNRAS.472.1618G} model. Nevertheless, in order not to underestimates the spike lifetimes, we multiply the values by a factor of two as suggested by \citealt{2021MNRAS.508..267S}. This factor was derived from simulations of solar spots, including various number of spots, inclination angles and rotation periods.
The spike lifetimes were obtained for all targets in our sample, and the results are in the HR diagram from Fig.\,\ref{fig:HRD_time_decay}, where these dictate the colour code. The spike lifetime values are also listed in Table \ref{tab:spike_param1}.

\begin{figure}
	\includegraphics[width=\columnwidth, scale=2]{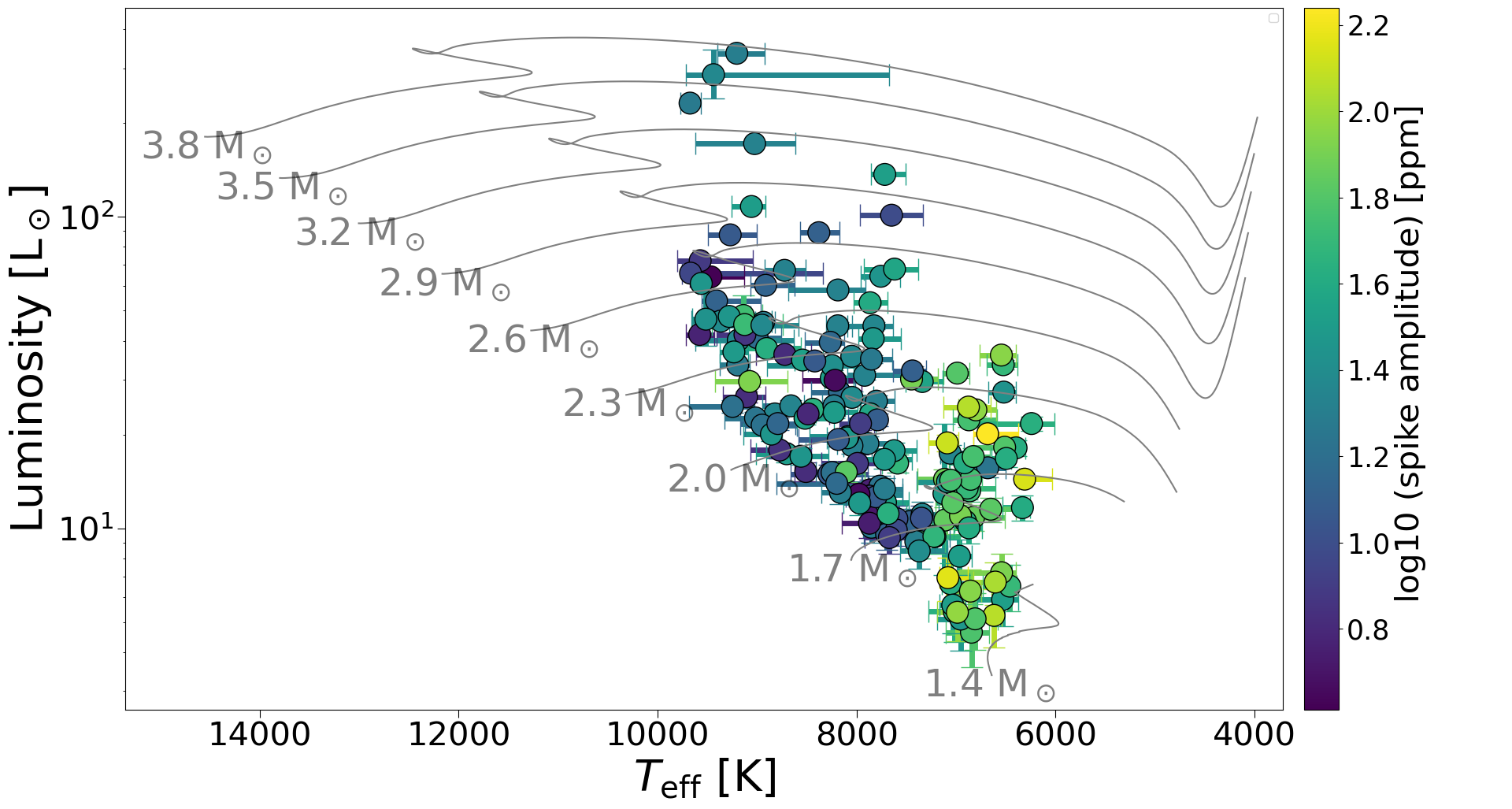}
    \caption{ HR diagram with the colour code illustrating the amplitude of the spikes. The source for luminosity and $T_{\rm eff}$ values is described in section \ref{sec:lum_teff}. Warszaw-New\,Jersey evolutionary tracks ($Z=0.012$, \citealt{2004A&A...417..751A}) are displayed in the background for guidance only.}
    \label{fig:HRD_spike_ampl}
    
\end{figure}

\section{Results}
\label{sec:results}

We analysed the time series of 162 \textit{hump \& spike} stars observed with the Kepler telescope and extracted the spike frequencies corresponding to the rotational frequencies, amplitude and corresponding uncertainties. We calculated the spike lifetimes, used stellar radius to determine the rotational velocity and used luminosity values from \citet{2019MNRAS.485.2380M}, \citet{2020AJ....159..280B} and \textit{Gaia} DR2 and $T_{\rm eff}$ values from \textit{Gaia} DR2 catalogue to place our targets in HR diagrams.

Fig.\,\ref{fig:frot_comparison} shows the difference between the obtained rotational frequency from this work and values from other sources mentioned in the caption and legend; most of our values in the literature agree with ours. We checked each star that does not have similar rotation frequency values to the ones previously reported and concluded that the values in this work are correctly identified as the rotation frequency. The few differences come from misidentifying the main spike with one of its harmonics. For example, KIC\,3868032 was reported by \citet{2013MNRAS.431.2240B, 2014MNRAS.441.3543B, 2017MNRAS.467.1830B} to have a rotation frequency of 0.4 $\mathrm{c d^{-1}}$. \citet{2020MNRAS.492.3143T} reported a frequency of 1.67 $\mathrm{c d^{-1}}$ which is in agreement with our value. A more thorough investigation, revealed that \citet{2020MNRAS.493.4518K} reported a  \textit{v} sin \textit{i} value of $181 \pm 8$ $\mathrm{km \ s^{-1}}$ for this star. A value for the rotational velocity using the 0.4 $\mathrm{c d^{-1}}$ rotational frequency and a radius value of $2.4^{+0.07}_{-0.06} R_{\odot}$  \citep{2020AJ....159..280B} would yield a rotational velocity of $\approx 49$ $\mathrm{km \ s^{-1}}$ . Given that the \textit{v} sin \textit{i} value is the lower limit for the rotational velocity, this value would contradict the result of \citet{2020MNRAS.493.4518K}. Assuming the rotation frequency of 1.67 $\mathrm{c d^{-1}}$, the rotational velocity would be $206^{+6}_{-5}$ $\mathrm{km \ s^{-1}}$,  which is consistent with the measured \textit{v} sin \textit{i}. A star can be represented more than once in Fig.\,\ref{fig:frot_comparison}, as stars have been reported more than once in previous studies.  

We explore different parameters for any significant trends. Fig.\,\ref{fig:corr_plots} shows the correlations between stellar parameters and various parameters extracted in our analysis. In \ref{fig:corr_plots}.a2, a moderate correlation between the rotational velocity and effective temperature is shown. This correlation is not surprising as it has been known for decades that hotter stars rotate faster than their cooler counterparts (e.g. \citealt{1978trs..book.....T}). A similar correlation can be seen in \ref{fig:corr_plots}.b2 between luminosity and rotational velocity, which is interesting. An explanation is that the more luminous stars in the sample are the evolved higher-mass stars, and the rapid rotation they had while closer to ZAMS still dominates over the spin-down as their radii increase with age.
Yet, we must highlight the bias induced by the target selection when discussing these correlations. Targets that originate from \citet{2020MNRAS.492.3143T}, are stars initially selected by \citet{2013MNRAS.431.2240B} to correspond roughly to spectral type A9-A0. The temperature range of interest was 7500 < $T_{\rm eff}$ < 10 000 K. This clear $T_{\rm eff}$ cut off is noticeable also in Fig.\,\ref{fig:clean_HR}, where stars symbolized by green circles, lie roughly in this temperature range (differences may occur due to the fact that \citet{2013MNRAS.431.2240B} used the $T_{\rm eff}$  from the KIC catalogue, while the values used in Fig.\,\ref{fig:clean_HR} were from \textit{Gaia} DR2). A similar trend, but maybe not as significant, can be observed in Fig.\,\ref{fig:corr_plots}.c2, where the rotational velocity is compared to the stellar radius. Considering all of the above and that more massive, hotter stars rotate faster (Fig.\,\ref{fig:HRD_vrot}), under the assumption that magnetic activity gives rise to the spike frequency, these results could also suggest that more massive stars require a faster rotation rate to generate observable effects in the form of \textit{hump \& spike}.

In addition to the initial sample of stars with $T_{\rm eff}$ = [6500, 10000]\,K, we also attempted to populate our target list with cooler stars by visually inspecting selected stars from \citet{2021ApJS..255...17S}. The latter study concentrates on G and late F main-sequence and cool subgiant stars (in Fig.\,\ref{fig:clean_HR}, stars symbolised with blue circles). Only three stars with $T_{\rm eff}$ = [6310, 7033]\,K were found, suggesting that the \textit{hump \& spike} feature does not occur in cooler stars. The reason may be that lower-mass stars rotate on average slower, and/or the stellar structure is different, but we stress that a more systematic search for \textit{hump \& spike} stars is necessary in order to suggest a possible explanation. However, in the OsC modes scenario we do not expect cooler \textit{hump \& spike} stars as the convective core does not exist at masses lower than roughly $1.3~\rm M_{\odot}$. Additionally the thick convective envelope of late type stars would not allow g~modes to penetrate all the way to the surface.

There seems to be no apparent connection between any parameter and the spike lifetime. The average spike lifetime value for the whole sample is around 30\,d, with $\sim$$95$ per cent of the values being less than 60\,d. The order of magnitude of the spike lifetime values are in agreement with values of starspot lifetime reported by \citet{2017MNRAS.472.1618G} regarding F-type stars. Although the F-stars in the \citet{2017MNRAS.472.1618G} sample are not very numerous, they report that the stellar spots do not survive very long (unlike in the case M, K and G stars) nor reach substantial sizes. 

The missing correlation between the spike lifetimes and stellar mass is hard to reconcile in the OsC scenario. As the g modes would be modulated by the convective turnover time scales, we expect the spike lifetimes to decrease with stellar mass. We address this in more detail in section \ref{sec:conv_core}.  

The weak correlations from Fig.\,\ref{fig:corr_plots}.a1, which shows the relation between the spike amplitude and $T_{\rm eff}$ and Fig.\,\ref{fig:corr_plots}.b1 that depicts the relation between the spike amplitude and luminosity (\ref{fig:corr_plots}), suggest that cooler and less luminous stars have higher spikes. The same trend clearly showing that more massive stars have lower spike amplitudes is illustrated in Fig.\,\ref{fig:HRD_spike_ampl}. This correlation is further explored and discussed in section \ref{sec:comp_magn}, as it is related to the predictions of \citetalias{Cantiello_2019}.

\section{Discussions}
\label{sec:discussions}

\subsection{Magnetic field strength}

\label{sec:mag_discussions}

\citet{Cantiello_2019} argue that the thin envelope convection zones of intermediate-mass stars can sustain dynamo-generated magnetic fields. 
If the observed photometric variability is due to periodic temperature variations caused by magnetic spots, one can estimate the magnetic field strength using the amplitude of the spike and some assumptions for the spot filling factor. 
Following \citet{2011A&A...534A.140C},
we start from the definition of the radiative gradient:
\begin{equation}
    \nabla_{\rm rad} = \left(\frac{dlnT}{dlnP}\right)_{\rm rad} \simeq \frac{P}{T}\frac{\Delta T}{\Delta P} ,
    \label{eq:rad_grad}
\end{equation}
which can be re-written as 
\begin{equation}
    \frac{\Delta T}{T} = \nabla_{\rm rad} \frac{\Delta P}{P}.\label{eq:rad_grad2}
\end{equation}
We then assume that a magnetic spot is in hydrostatic and thermal equilibrium with its surroundings. The latter requires  the radius of the spot to be significantly smaller than the stellar radius. We also assume that the radiative gradient is not affected substantially by the presence of the magnetic field. 
Assuming hydrostatic equilibrium implies that a magnetic spot has a smaller gas pressure compared to its non-magnetized surroundings, since part of the total pressure is provided by the magnetic field. Together with the assumption of thermal equilibrium, this implies a lower density in the magnetized region. A local depression in the surface density allows photons to emerge from deeper regions of the star, so an observer looking down into a magnetic spot at the surface of a radiative star is expected to see hotter temperatures compare to a non-magnetized region \citep{2011A&A...534A.140C}. Using Eq.~\ref{eq:rad_grad2} this temperature contrast can be written as
\begin{equation}
    \frac{\Delta T}{T} = \nabla_{\rm rad} \frac{P_{\rm mag}}{P_{\rm tot}} = \frac{\nabla_{\rm rad}}{\beta}  
\end{equation} 
where 
\begin{equation*}
\beta = \frac{P_{\rm tot}}{P_{\rm mag}} = \frac{P_{\rm tot}}{B^2/8\pi},
\end{equation*}
and $P_{\rm mag}$ is the component of the pressure due to the presence of the magnetic fields, $P_{\rm tot}$ is the total pressure and $B$ the magnetic field strength. Assuming that the ratio between the magnetic pressure $P_{\rm mag}$ and the total pressure  $P_{\rm tot}$ (from the gas and radiation) is very small, $\beta \gg 1$,  and using $ L = 4 \pi R^2 \sigma T_{\rm eff}^4 $, (where R is the stellar radius), we can obtain a relation to estimate the local contrast in luminosity between the spot and the rest of the stellar surface \citep{2011A&A...534A.140C}:
\begin{equation}
\centering
\frac{\Delta L}{L} \simeq \frac{4 \Delta T}{T} = \frac{4\nabla_{\rm rad}}{\beta}
\label{eq:Lum_ratio}
\end{equation}

This equation provides the luminosity contrast of a magnetic spot compared to the unperturbed stellar luminosity. To compare this to observations, one has to take into account the filling factor  $f=(r/R)^2$, where $r$ is the radius of the spot and R is the stellar radius. Note the degeneracy between temperature (or luminosity) contrast and filling factor: A large spot with a small luminosity contrast can give a similar brightness variation to a smaller spot with a higher luminosity contrast. From a theoretical standpoint, \citealt{2011A&A...534A.140C} and \citetalias{Cantiello_2019} suggested that magnetic fields in hot stars could produce features on scales that have comparable sizes with (or larger than) the pressure scale height of the convective regions where they are generated. Lacking firm observational constraints on the size of the spots that might be generating the brightness fluctuations, we will assume the spot size $r$ to be comparable to the pressure scale height in sub-surface convective layers. As we will discuss below the spots could actually be larger, so this assumption means our estimates for the magnetic fields represent upper limits.  Equation \ref{eq:Lum_ratio} becomes:

\begin{equation}
\frac{\Delta L}{L} =  \frac{4 \nabla_{\rm rad}}{\beta} \left(\frac{H_{p}}{R}\right)^2
\label{eq:Lum_ratio_geom}
\end{equation}

\noindent where $H_p$ is the pressure scale height. Using the definition of $\beta$, equation \ref{eq:Lum_ratio_geom} can be written as:

\begin{equation}
\frac{\Delta L}{L} =  \frac{4 \nabla_{\rm rad} B^2}{8 \pi P_{\rm tot} } \left(\frac{H_p}{R}\right)^2
\label{eq:Lum_ratio_geom3}
\end{equation}

\noindent The ratio $\Delta L/L$ is then replaced with the observed photometric amplitude ($A_{\rm spike}$) in the Kepler data. Note that our calculation assumes a single spot and we do not correct for the \textit{Kepler} bandpass. This is not necessarily correct, and the data suggest some stars have more than one spot, see e.g. the phase plots in Section \ref{sec:phase_plots}.  The magnetic fields amplitude is then recovered using the following relation:

\begin{equation}
    B^2 \simeq A_{\rm spike} \frac{ 2 \pi P_{\rm tot} }{\nabla_{\rm rad}} \left(\frac{R}{H_p}\right)^2.
    \label{eq:B_squared}
\end{equation}

From a grid of ten stellar models in the mass range $1.25-4\, {\rm M}_{\odot}$, obtained using MESA \citep{2011ApJS..192....3P, 2013ApJS..208....4P, 2015ApJS..220...15P, 2018ApJS..234...34P, 2019ApJS..243...10P}, we have extracted through linear interpolation values for the total pressure at the surface $P_{tot}$, radiative gradient $\nabla_{rad}$ and the pressure scale height $H_p$. The MESA models are accessible through github \footnote[1]{\url{https://github.com/matteocantiello/hump_spikes_stars}}. The stellar models have solar metallicity ($Z=0.02$) and a chemical composition mixture taken from \citet{2005ASPC..336...25A}. The mixing length parameter was assumed to be 1.6. The $P_{tot}$ values we obtain for our sample span from 375 to 17497 Pa, with a median value of 3092 Pa. The average radiative gradient values for our stars is 0.126, spanning from 0.125 to 0.132. An average value for the pressure scale height for the stars in our sample was found to be $0.01 R_{\odot}$, ranging from 0.005 to 0.02 $R_{\odot}$.

We note that the $H_p$ values are the lower limits to the sizes of the magnetic features. \citetalias{Cantiello_2019} pointed out that magnetic features expand when rising towards the stellar surface depending on the ratio of the density in the convective layers and at the photosphere. Moreover, intermediate-mass stars tend to be rapidly rotating, which can lead to dynamo action on scales larger than the local pressure scale-height \citep{Brandenburg:2005,2011A&A...534A.140C}. Therefore the recovered values of $B$ are upper limits.

\begin{figure}
	\includegraphics[width=\columnwidth, height=135pt]{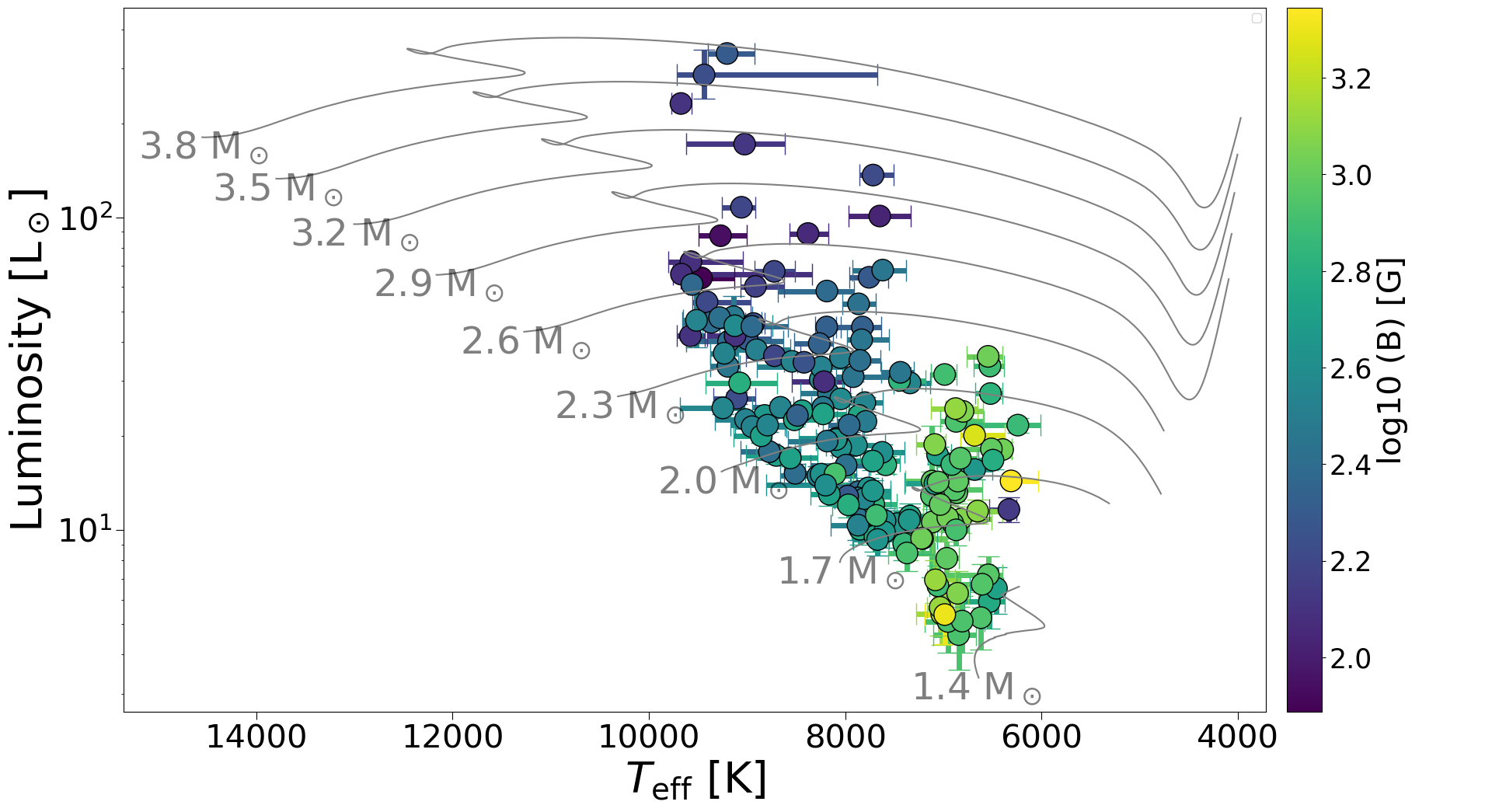}
    \caption{ HR diagram with colour code illustrating the estimated magnetic field strengths. The source for luminosity and $T_{\rm eff}$ values is described in section \ref{sec:lum_teff}. Warszaw-New\,Jersey evolutionary tracks ($Z=0.012$, \citealt{2004A&A...417..751A}) are displayed in the background for guidance only.}
    \label{fig:HRD_B_strength}
\end{figure}

\begin{figure}
	\includegraphics[width=\columnwidth, height=135pt]{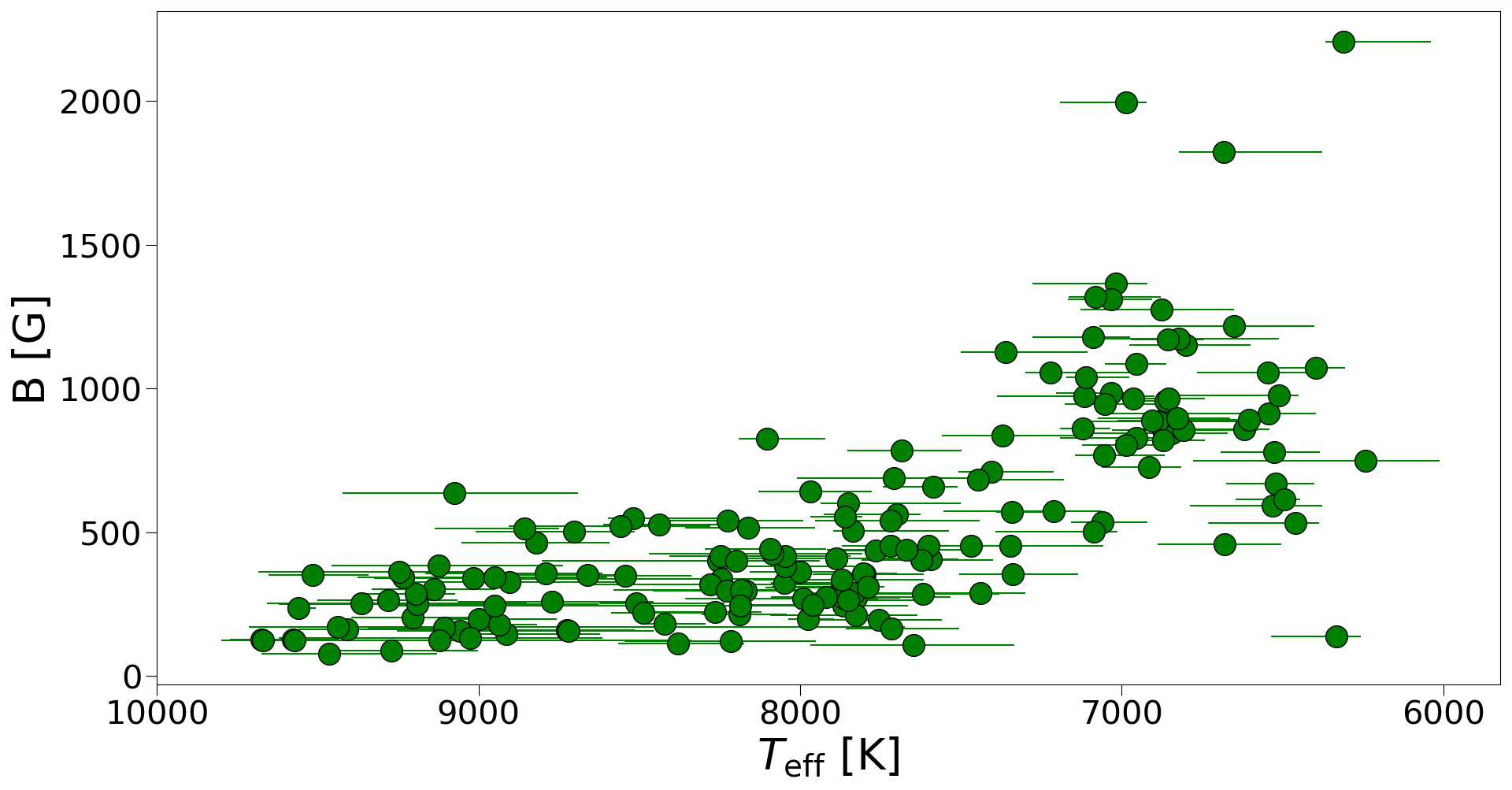}
    \caption{ Estimate values for the magnetic field strength with respect to the $T_{\rm eff}$ of the stars.}
    \label{fig:B_vs_teff}
\end{figure}

Under all the assumptions made above we computed upper limits estimates for the magnetic field strength in our stars sample. The results are depicted in Fig.\,\ref{fig:HRD_B_strength}. The obtained values are also shown with respect to $T_{\rm eff}$ in Fig.\,\ref{fig:B_vs_teff} and range from 77 to 2207\,G, with an average value of 540\,G. In the next section we discuss our findings in the context of theoretical predictions.

\subsection{Comparison with predictions from literature}
\label{sec:comp_magn}

Fig.\,\ref{fig:HRD_spike_ampl} clearly shows that the spike amplitude is decreasing with increasing stellar mass, which in the context of magnetic fields, suggests stronger magnetic fields for lower-mass stars, which also have deeper (sub)surface convective layers. 

Based on the spike amplitude we have estimated the strengths of equipartition magnetic fields in the convective envelopes. We note again that the intrinsic values of the amplitudes depend on e.g., stellar inclination and spot latitudes, which are unknown. Nevertheless, our results are in agreement with the predicted values of average equipartition magnetic fields in the sub-surface convective zones (see figure 6 from \citetalias{Cantiello_2019}.) By comparing Fig. \ref{fig:HRD_B_strength} and figure 8 from \citetalias{Cantiello_2019}, a similar general trend can be seen. Both the predicted magnetic field strengths in \citetalias{Cantiello_2019} and the calculated values from this work decrease with increasing stellar mass. Furthermore, Fig. \ref{fig:HRD_B_strength} shows that more evolved stars have stronger calculated magnetic fields, which also agrees with the predictions from \citetalias{Cantiello_2019}, as the convective envelope gets deeper as the star evolves. 

The strengths of the magnetic fields at the surface that were computed by \citetalias{Cantiello_2019} were determined by scaling the equipartition fields with the density inside the HeII convective zone, ranging from 2000\,G to less than 1\,G (see figure 9 from \citetalias{Cantiello_2019}).  Their findings showed that for a star like Vega the expected mean longitudinal field  would be very small ($\sim\!10^{-2} \,\rm G$), which is consistent with the observed value of $0.6 \pm 0.3 \rm\, G$ \citep{2009A&A...500L..41L}. Therefore the values obtained with the density scaling relation \citetalias{Cantiello_2019} proposed should be taken as a lower limit. This also brings forth the fact that the magnetic buoyancy might not necessarily be the process that brings the magnetic field to the surface. Other processes have been proposed such as the one proposed by \citet{2010A&A...523A..19W,2011A&A...534A..11W}, where the magnetic field would reach the surface through magnetic tension, in which case the strength at the surface could be larger, comparable to the equipartition values. 

The calculated values for the spike lifetimes (Fig. \ref{fig:HRD_time_decay}) suggest that if present, the magnetic features are not very long lived, having lifetimes of tens of days, with only five stars surpassing 60\,d. This is in agreement with predictions from \citetalias{Cantiello_2019}, regarding the rapidly evolving features of the dynamo-generated magnetic fields. 
In addition, the trend that can be observed for the spike amplitude in Fig.\,\ref{fig:HRD_spike_ampl} is not present for the spike lifetime. This suggests that a more complex process may determine the spike lifetime.

To summarize, the spike amplitude trend with mass and $ T_{\rm eff}$ in the context of magnetic fields, suggests that the dynamo in the subsurface convection hypothesis is favoured over the `failed-fossil' scenario. 

\subsection{Convective core rotation}
\label{sec:conv_core}

\subsubsection{Convective turnover time}

As described above, \citet{LeeSaio2020} and \citet{Lee2021} suggest that Overstable convective (OsC) modes could, in the presence of radial differential rotation, couple with low-frequency g~modes in the envelope to emerge on the surface. A 20\% differential rotation is assumed in \citet{LeeSaio2020}, while \citet{Lee2021} assumes the core rotation to be 10\% higher than the envelope rotation.

An OsC mode of an azimuthal order $m$ would have a frequency $m$ times the core rotation frequency. Here we investigate whether the rotational modulation characterised by the spike could be due to OsC modes. 

If the spikes corresponded to $m$ times the convective core rotation rates, it would mean that the spike lifetimes calculated in subsection \ref{sec:time_decay} would be associated with the movement of the convective cells inside the core. In other words, we would expect the spike lifetimes to be related to the core's convective turnover time. It is well established that the mass of the convective core increases with stellar mass (see, e.g., figure 22.7 in \citealt{1990sse..book.....K}). Consequently, the spike lifetimes calculated in section \ref{sec:time_decay} should correlate with the stellar mass. 
From MESA models, we have extracted a few values for the convective turn overtime in the stellar mass regime of interest. These values were obtained from models which have solar metallicity ($Z=0.02$) and a chemical composition mixture taken from \citet{2005ASPC..336...25A}. The mixing length parameter was assumed to be 1.6 (Table \ref{tab:conv_oveturn_times}). As can be seen in Table \ref{tab:conv_oveturn_times} the convective turnover times decrease with stellar mass. This negative correlation is expected to be present throughout the entire main-sequence evolutionary stage. 

\begin{table}
	\caption{Extracted values of convective turnover time from MESA models for a few representative stellar masses.}

	\label{tab:conv_oveturn_times}
	\begin{tabular}{|cc|cc|} 
		\hline

	    \textbf{ZAMS}& &\textbf{TAMS}& \\
	    \hline
		Mass  & $\tau_{\rm turn}$ & Mass & $\tau_{\rm turn}$   \\ 
  
		[${\rm M}_{\odot}$] &[days] &[${\rm M}_{\odot}$] &[days]  \\ 
		\hline
		
        1.5&173&1.5&44 \\
        2.0&113&2.0&38 \\
        2.4&83&2.4&35 \\
        3.0&68&3.0&32 \\
        3.6&62&3.6&30 \\
        4.0&59&4.0&29 \\
		\hline

        \multicolumn{4}{l}{ $\tau_{\rm turn}$ = convective turnover time}\\

	\hline
	\end{tabular}
\end{table}

It is important to point out that our results do not show any correlation between the spike lifetime and stellar mass, as seen in the HR diagram from Fig.\,\ref{fig:HRD_time_decay}, suggesting that the broadness of the spike is not connected to convective turnover time in the core. 

In the context of OsC modes, a possible explanations for the observed trend in amplitude seen in Fig.\,\ref{fig:HRD_spike_ampl}, could be the changing density of g~modes with stellar mass. In other words, the amplitude of an OsC mode at a surface should be larger when the resonance with a g~mode is stronger, which would occur more frequently if the density of g~modes (in the co-rotating) frame is higher. This would take place in less massive stars and could imply that the amplitudes of the spikes tend to be lower in more massive stars, as seen in Fig. \ref{fig:HRD_spike_ampl}.

\subsubsection{Harmonic signature and rotation}
\label{sec:harmonic_rotation}

\begin{figure}
	\includegraphics[width=\columnwidth]{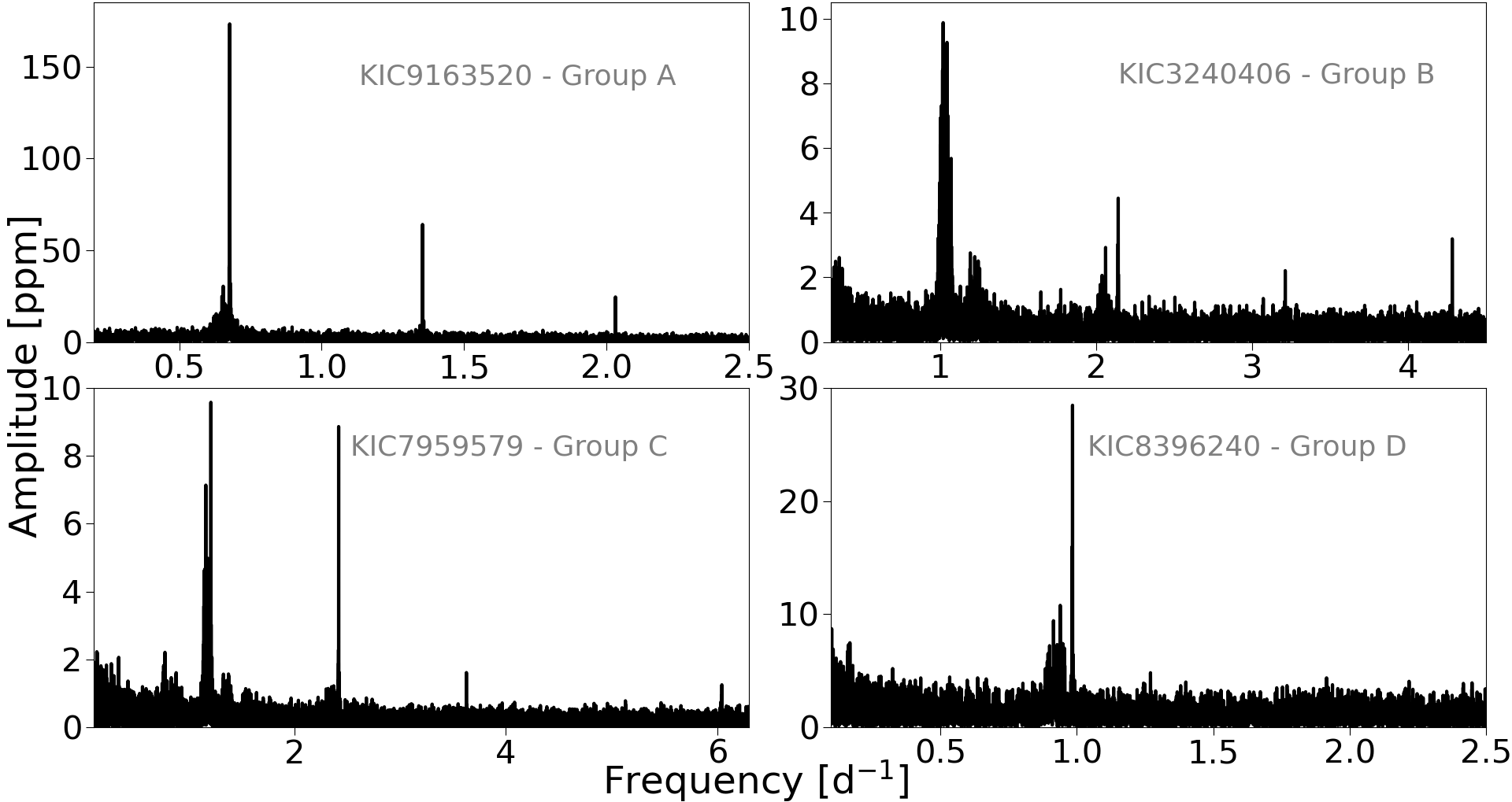}
    \caption{Harmonics signature of our sample. Group A: the amplitude of the spike harmonics decreases with increasing harmonic order. Group B: the amplitude of a harmonic is higher than a harmonic of a lower degree or the spike. Group C: harmonics are not consecutive, i.e. missing harmonics. Group D: no harmonics present. As described in the text in more detail Group B and C imply the presence of stellar spots, whereas Group A and D do not allow us to distinguish between spots and OsC modes.  }
    \label{fig:harmonic_sign}
\end{figure}

As mentioned in section \ref{sec:harm_spike}, only 14 stars do not exhibit detectable harmonics of the main spike (see an example in lower right corner of Fig.\,\ref{fig:harmonic_sign}). As seen in Fig.\,\ref{fig:hrd_harm}, the rest show harmonics in their Fourier spectra, with KIC\,7175896 having the highest amount of harmonics (8), with the highest harmonic being at 17 times the spike frequency. 

\citealt{LeeSaio2020} suggested that the expected amplitudes of g~modes resonantly excited by OsC modes would decrease as the azimuthal order ($m$) increases.  The variability induced is expected to be sinusoidal and would not cause any additional harmonics in the Fourier spectrum. In order to quantify the number of stars that would be consistent with this scenario, we subdivided our sample into four groups: \begin{itemize}
    \item Group A: the amplitudes of the spike and its harmonics decrease as the degree of the harmonic increases (see example in the upper left corner of Fig.\,\ref{fig:harmonic_sign}). This group comprises 111 stars and is consistent with both scenarios.
    \item Group B: a higher order harmonic has a higher amplitude than the previous harmonic or the main spike (upper right corner of Fig.\,\ref{fig:harmonic_sign}), which can only be explained by non-sinusoidal signals, such as spots. KIC\,4921184 (Fig.\,\ref{fig:kic4921184_fourier}) also falls under this category comprising 22 stars in total.
    \item Group C: the detected harmonic orders are not consecutive, i.e. there are missing harmonics, which again is an indication of non-sinusoidal signals, that could be induced by stellar spots. This group consists of 14 stars. The lower left corner of Fig.\,\ref{fig:harmonic_sign} shows an example where the fourth harmonic is not present while the third and fifth have similar amplitudes. 

    \item Group D: no harmonics detected, only the main spike is significant, not allowing us to draw any conclusion. We find 14 stars falling in this category. 
\end{itemize}

Counting only the harmonic behaviour or the absence of harmonics (Group A and D), we find 125 stars for which we cannot strictly determine the origin of the spike. In other words, both the OsC modes and the stellar spots caused by dynamo-generated magnetic fields could explain the spike feature. However, groups B and C clearly display a a behaviour that is compatible with stellar spots, making it a total of 36 stars.

We highlight that the OsC modes scenario, as suggested by \citet{LeeSaio2020} and \citet{Lee2021}, requires rapid rotation. As seen in Fig. \ref{fig:hist_frot_vrot} and \ref{fig:HRD_vrot} more than half of the stars from our sample have a rotational velocity higher than $100~\rm km\,s^{-1}$. However, the exact value for the required minimum rotation speed depends on model parameters such as the degree of radial differential rotation and the super-adiabatic temperature gradient in the rotating convective core.

The evolutionary stage of a star, i.e., whether it is still on the main sequence or not, has crucial implications for the OsC scenario. This is because our stars beyond the TAMS do not have a convective core. Based on their location in the HR diagram (e.g., Fig. \ref{fig:HRD_vrot}), we visually identified 34 stars that may have left the main-sequence. However, we note the following shortcomings to this identification:

    \begin{itemize}
    \item The visual identification was done without taking into account uncertainties in  $T_{\rm eff}$ and luminosity.
    \item The evolutionary tracks depicted in our HR diagrams do not take into account rotation and core overshooting and only use solar metallicity. This means that without detailed stellar modelling, determining whether a star still has a convective core is imprecise. 
    \item The effect of gravity darkening increases with rotation . Figure 4 from \cite{2014A&A...566A..21G} and figure 38 from \cite{2019ApJS..243...10P} nicely illustrate how the observed luminosity depends on the inclination angle for stars rotating with more than 50-60 per cent of their critical velocity. For these stars, the luminosity could be underestimated if observed pole-on or underestimated at the equator. For example, the two fast rotators in Fig. \ref{fig:HRD_vrot}, which lie around the $3.5~\rm M_{\odot}$ evolutionary track, could still be on the main sequence. 
 
\end{itemize}

\subsubsection{Phase-folded light curve}
\label{sec:phase_plots}

\begin{figure}

	\includegraphics[width=\columnwidth, height=135pt]{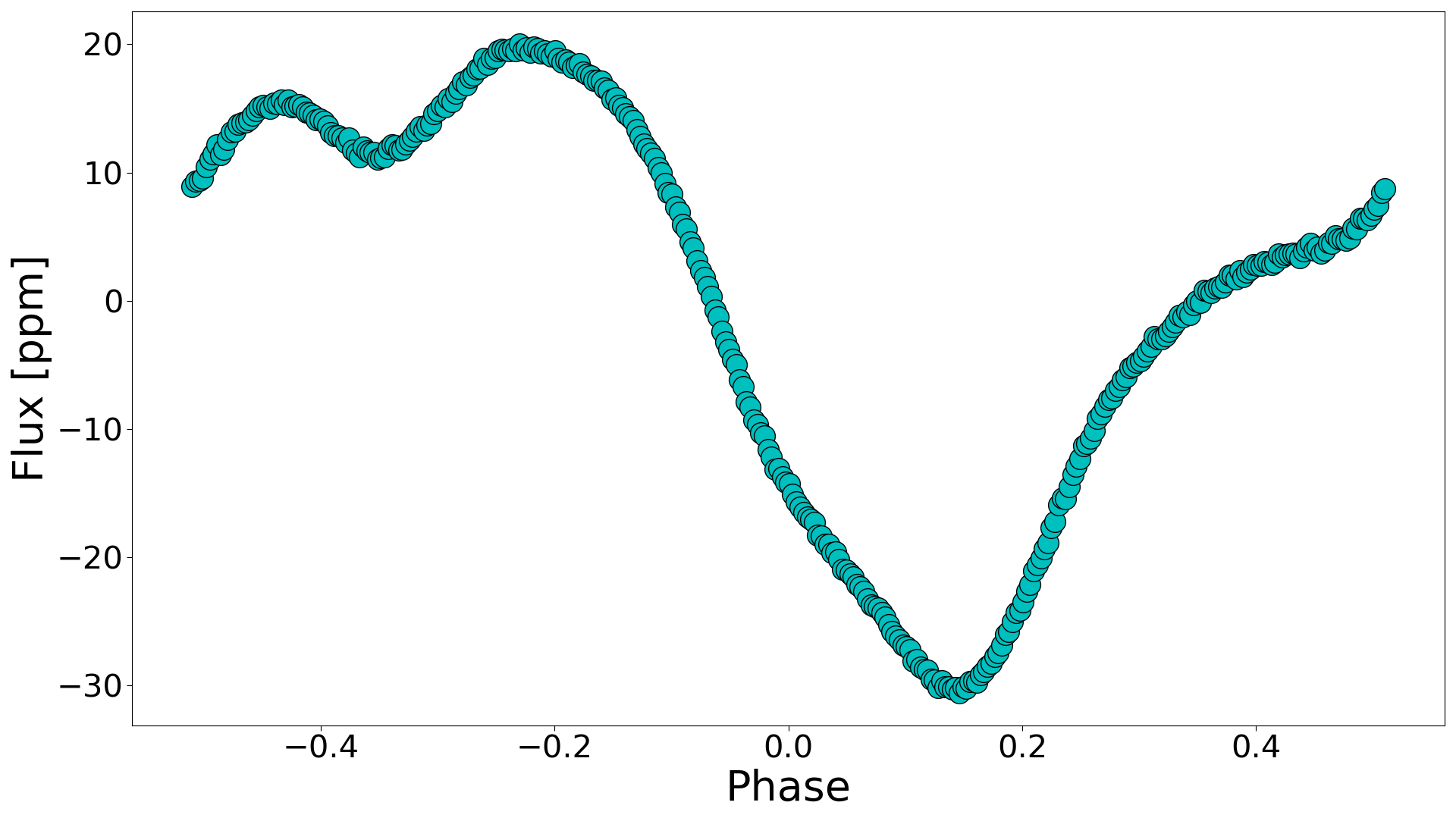}
    \caption{Phase-folded light curve computed with the band-passed filtered time series of KIC\,4661914 (group C - see section \ref{sec:harmonic_rotation}); average bin size $\sim$0.003 in phase folded on period  $\sim$$1.024$\,d which corresponds to the frequency of the main spike $\sim$$0.977$ $\mathrm{d^{-1}}$. }
    \label{fig:KIC4661914_phase_plot1}
\end{figure}

\begin{figure}

	\includegraphics[width=\columnwidth]{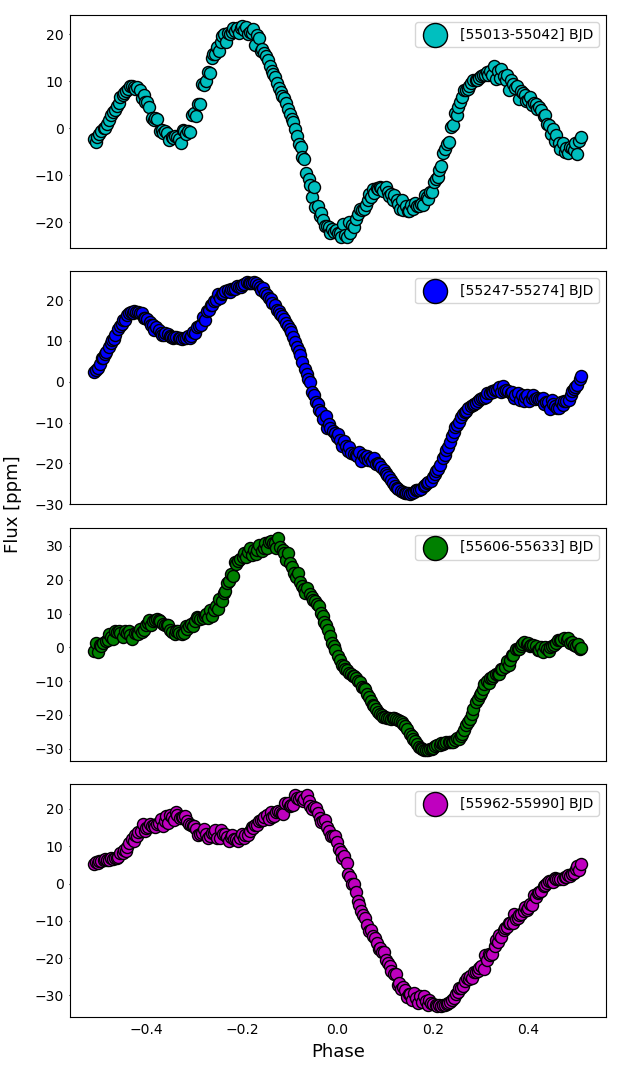}
    \caption{Phase-folded light curve computed with subsets of the band-passed filtered time series - KIC\,4661914 (group C - see section \ref{sec:harmonic_rotation}); average bin size for each plot $\sim$0.003 in phase folded on period  $\sim$$1.024$\,d (frequency of the main spike $\sim$$0.977 \mathrm{d^{-1}}$). The time range is noted for each plot; the epoch time is the same as in Fig.\,\ref{fig:KIC4661914_phase_plot1} (time of the first Kepler observation for KIC$\,$4661914).}
    \label{fig:KIC4661914_phase_plot2}
\end{figure}

\begin{figure}

	\includegraphics[width=\columnwidth, height=135pt]{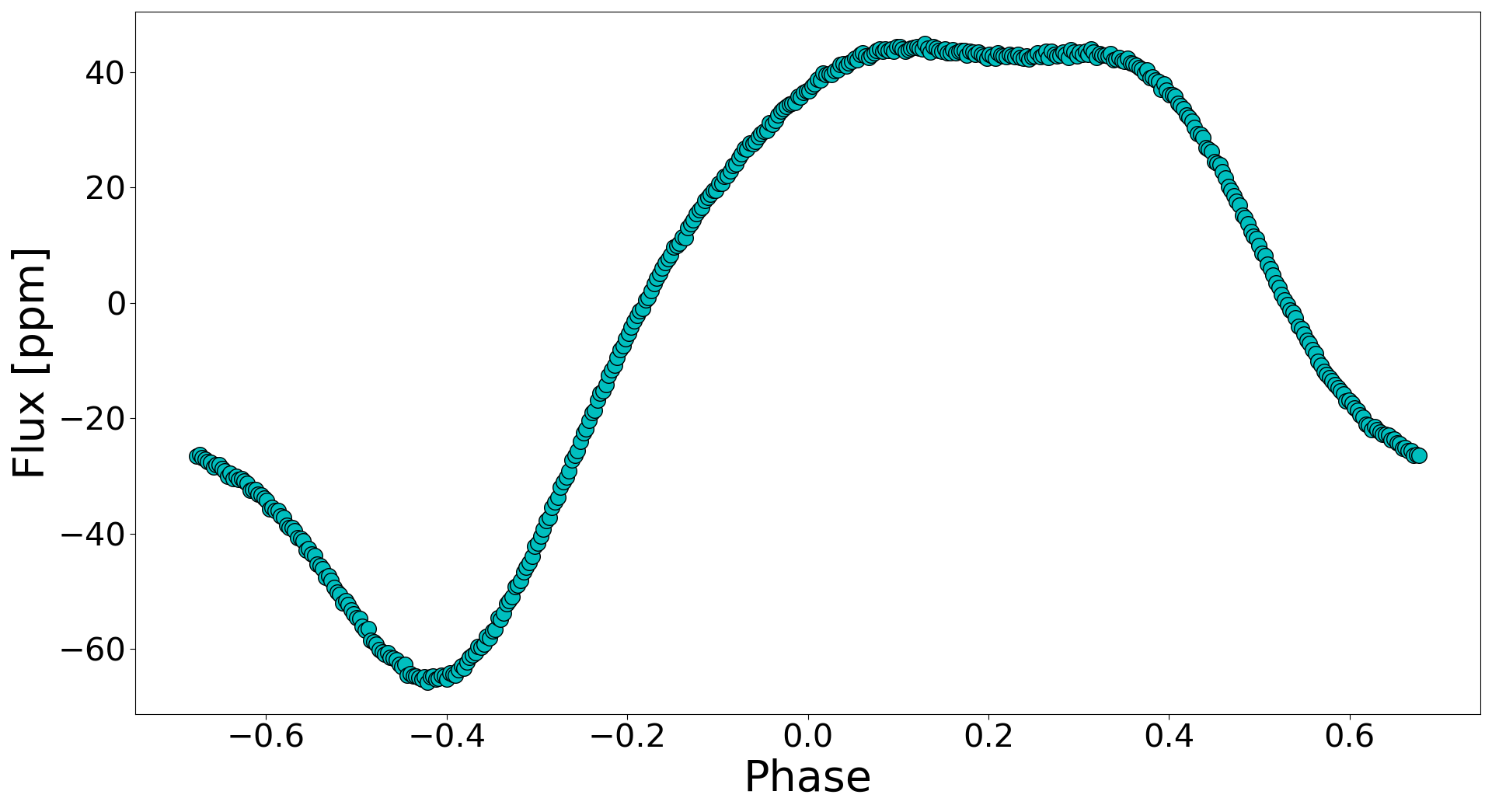}
    \caption{Phase-folded light curve computed with the full Kepler data - KIC\,2157489 (group B - see section \ref{sec:harmonic_rotation}); average bin size $\sim$0.003 in phase; folded on period  $\sim$$1.358$\,d; which corresponds to the frequency of the main spike $\sim$$0.736$\,$\mathrm{d^{-1}}$.}
    \label{fig:KIC2157489_phase_plot1}
\end{figure}

\begin{figure}

	\includegraphics[width=\columnwidth]{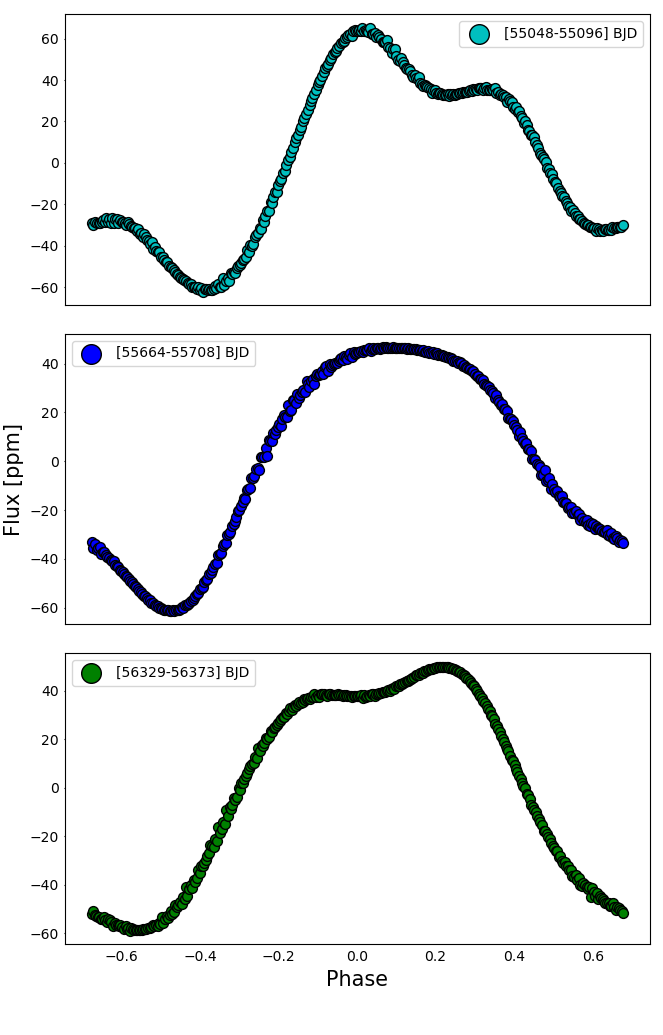}
    \caption{Phase-folded light curve computed with subsets of the full time series - KIC\,2157489 (group B - see section \ref{sec:harmonic_rotation}); average bin size $\sim$0.003 in phase; folded on period  $\sim$$1.358$\,d (frequency of the main spike $\sim$$0.736$ $\mathrm{d^{-1}}$); the time range is noted for each plot; the epoch time is the same as in Fig.\,\ref{fig:KIC2157489_phase_plot1} (time of the first Kepler observation for KIC\,2157489).}
    \label{fig:KIC2157489_phase_plot2}
\end{figure}

\begin{figure}

	\includegraphics[width=\columnwidth, height=135pt]{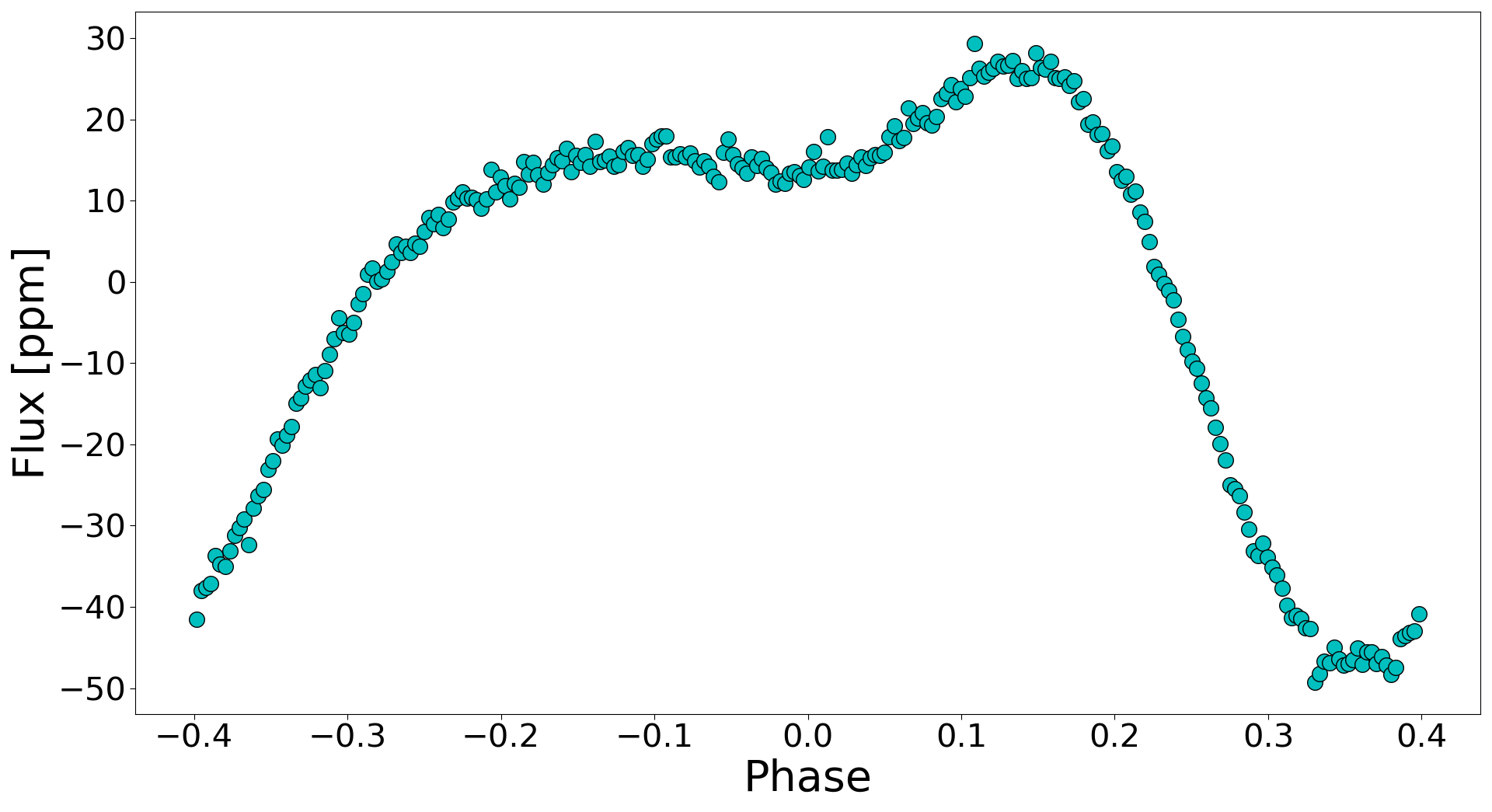}
    \caption{Phase-folded light curve computed with the full Kepler time series - KIC\,3440710 (group A - see section \ref{sec:harmonic_rotation}); average bin size $\sim$0.003 in phase; folded on period  $\sim$$0.8$\,d; which corresponds to the frequency of the main spike $\sim$$1.25$ $\mathrm{d^{-1}}$. }
    \label{fig:KIC3440710_phase_plot1}
\end{figure}

\begin{figure}

	\includegraphics[width=\columnwidth]{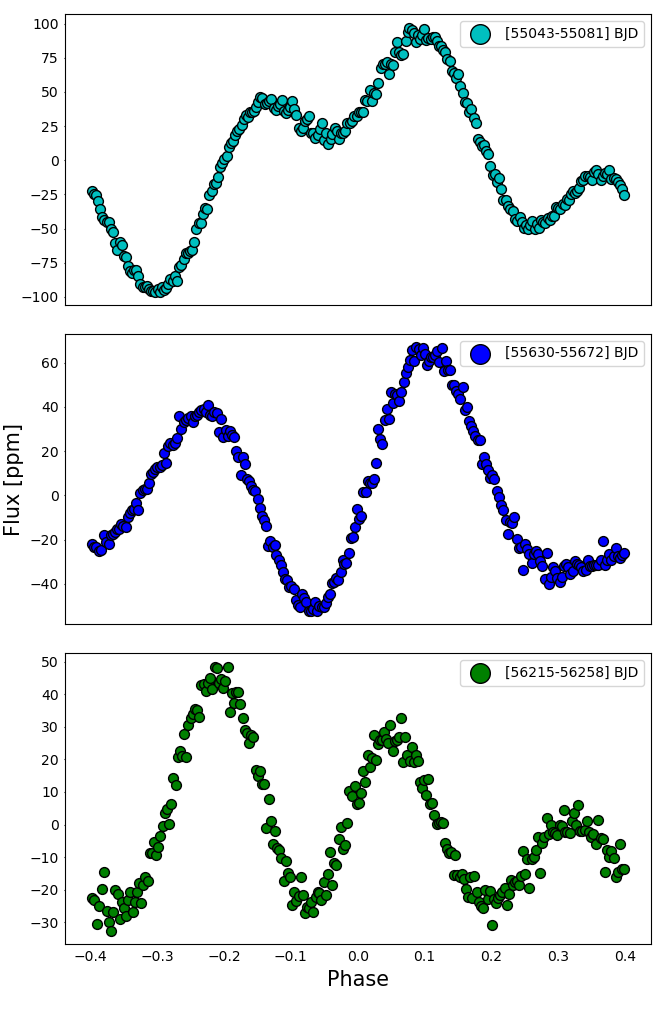}
    \caption{Phase-folded light curve computed with subsets of the full time series - KIC\,3440710 (group A - see section \ref{sec:harmonic_rotation}); average bin size $\sim$0.003 in phase; folded on period  $\sim$$0.8$\,d (frequency of the main spike $\sim$$1.25$ $\mathrm{d^{-1}}$); the time range is noted for each plot; the epoch time is the same as in Fig.\,\ref{fig:KIC3440710_phase_plot1} (time of the first Kepler observation for KIC\,3440710).}
    \label{fig:KIC3440710_phase_plot2}
\end{figure}

Another piece of evidence that might point towards the spike being associated with the stellar surface rotation rather than the convective core rotation, is the shape of the phase-folded light curves. Figures \ref{fig:KIC4661914_phase_plot1} - \ref{fig:KIC3440710_phase_plot2} depict the phase-folded time series of 3 stars from our sample. The groups defined in section \ref{sec:harmonic_rotation} to which these stars belong to are: KIC\,2157489 - B, KIC\,4661914 - C, KIC\,3440710 - A. All the data were folded with the periods derived from the spikes. The phase folded light curves were obtained with \textit{lightkurve} \citep{2018ascl.soft12013L} and afterwards binned in phase with the values mentioned in the captions. Figures \ref{fig:KIC4661914_phase_plot1}, \ref{fig:KIC2157489_phase_plot1} and \ref{fig:KIC3440710_phase_plot1} represent the phase-folded light curves of the full Kepler data sets. We note that if the median amplitude value is taken in a given bin, the effect of amplitude change with time is not visible when displaying the entire data set. Figures \ref{fig:KIC4661914_phase_plot2}, \ref{fig:KIC2157489_phase_plot2} and \ref{fig:KIC3440710_phase_plot2} contain the phase-folded light curves computed with sub-data sets. It is clear that the wiggly shape of the light curve changes depending on the time range of the sub-data set.   Our phase-folded data show a characteristic non-sinusoidal signal that could be explained by the presence of several evolving stellar spots.

\section{Conclusions and future work}
\label{sec:conclusion}

In this work, we analysed the Kepler light curves of 213 stars that present a similar feature in their Fourier spectra called \textit{hump \& spike}. Out of this sample, we have selected 162 stars that did not present obvious photometric signs of being in a binary system (e.g. no transits or ellipsoidal variation). We aimed to validate the initial theoretical interpretation of the origin of the spike \citep{2018MNRAS.474.2774S}, which is assumed to be rotational modulation induced by stellar spots co-rotating with the stellar surface. We tested the theory regarding the presence, origin and strength of magnetic fields on our sample of stars under the assumption that the hump represents unresolved Rossby modes and the spike stellar spots. In the present work, we have concentrated only on the spike in the \textit{hump \& spike} feature. The 'hump' will be the focus of future work currently in preparation.

In addition, we have also discussed another possible phenomenon that could cause rotational modulation in the light curves of our stars, as suggested by \citet{LeeSaio2020} and \citet{Lee2021}. In this context, the spike would be due to Overstable Convective (OsC) modes that resonantly excite g~modes in the co-rotating frame propagating to the surface and consequently become observable. The spike frequency in this scenario would correspond to the convective core rotation frequency. 

We have extracted the spike frequency and amplitude for all stars and used stellar parameters such as $T_{\rm eff}$, luminosity and the stellar radius from various sources (see Table \ref{tab:spike_param1} to consult the source of the stellar parameters). We determined the rotational velocities using stellar radii and spike frequencies. Further, we determined the lifetime of the spike using autocorrelation functions and extracted the number of spike harmonics present in the Fourier spectra of our stars. In favour of the spike being evidence for stellar spots due to magnetic fields, we find the following:
\begin{itemize}
    \item The spike amplitudes are higher for cooler stars and also slightly higher for evolved stars, mirroring predictions regarding the strength of dynamo-generated magnetic fields in (sub) surface convective layers.
    \item The spike lifetimes are short, of the order of tens of days, similar to what one would expect from magnetic features of dynamo-generated magnetic fields.
    \item The spike lifetimes do not correlate with stellar mass. This would exclude the possibility that the spike frequency would correspond to the convective core rotation frequency, as it should be modulated by the convective turnover time, which increases with stellar mass. 
    \item The harmonic signature of 36 stars suggests a non-sinusoidal signal consistent with co-rotating stellar spots.
    \item Phase-folded light curves of time series containing the signal induced by the spike and its harmonics suggest a non-sinusoidal signal specific to changing stellar spots.
   \item The estimated magnetic field strengths, however uncertain, are of the same order of magnitude as predictions of non-Ap stars. 
   \item We find 34 stars that could be in a post main-sequence evolutionary stage, which excludes the OsC modes scenario, as these stars do not have a convective core. We note, however, that this identification is visual only. The evolutionary tracks used here do not take into account, e.g., core overshooting, rotation and non-solar metallicities. In addition, the luminosity values used here are not corrected for the gravity-darkening effect, which is considerable in the case of high stellar rotation.
\end{itemize}

Our results suggest that, despite their very shallow convective envelopes, these stars could have spots likely induced by magnetic fields. The characteristics of the features causing the brightness variability in the shape of the spike point towards the idea of a dynamo-generated magnetic field as suggested by \citetalias{Cantiello_2019}. The short spike lifetimes, the estimated magnetic field strengths and presumed magnetic feature sizes favour dynamo-generated fields rather than fossil fields.  
While these stars are unfortunately too faint to qualify for spectropolarimetry, recent observations of $\beta$ Cas - a rapidly rotating $\delta$ Scuti star - indicate that dynamo-generated magnetic fields exist in intermediate mass stars \citep{2020A&A...643A.110Z}.  
As observations are biased towards stars with strong magnetic fields and large magnetic features, our work provides a path for future observational efforts towards weaker magnetic fields that generate small-scale features. 

Furthermore, ongoing observational efforts from the ground and the TESS satellite will allow us to determine whether the spots (if present) in the \textit{hump \& spike} stars are bright spots, as suggested by \citetalias{Cantiello_2019}. Our work also offers the chance to verify and test the predictions and theory of \citet{2018MNRAS.474.2774S}, where it was hypothesised that Rossby modes are mechanically excited by deviated flows caused by stellar spots. An obvious next step is to find bright \textit{hump \& spike} stars observed with TESS, which will allow us to measure the magnetic fields directly using ground-based facilities.

In favour of the spike being evidence of g~modes being resonantly excited by OsC modes from the convective core we find the following arguments:
\begin{itemize}
    \item The spike amplitude increasing with mass could be because the amplitude of OsC modes is larger when the resonance with g~modes is stronger. In the case of lower mass stars, the density of g-mode frequencies is higher; therefore, it would be more likely that the resonance of g~modes in the co-rotating frame is stronger.
    \item The harmonic signature (amplitude decreases as azimuthal order decreases) in the case of 111 stars, where the OsC modes scenario could be the cause for the rotational modulation. This applies also for 14 stars more which do not exhibit detectable harmonics of the main spike.
\end{itemize}

Based on our analyses and sample of stars, neither scenario can be entirely excluded at this stage. It is also possible, although rather unlikely, that both stellar spots and OsC modes could generate the \textit{hump \& spike} feature. While in this work, we find more arguments supporting the idea of stellar spots induced by magnetic fields, as summarised above, a full conclusion can be reached only by directly measuring magnetic fields in \textit{hump \& spike} stars.

\section*{Acknowledgements}
This work was supported by a research grant (00028173) from VILLUM FONDEN. Funding for the Stellar Astrophysics Centre is provided by The Danish National Research Foundation (Grant agreement no.: DNRF106). This research was supported in part by the National Science Foundation under Grant No. NSF PHY-1748958.

S.M.~acknowledges support from the Spanish Ministry of Science and Innovation (MICINN) with the Ram\'on y Cajal fellowship no.~RYC-2015-17697 and grant no.~PID2019-107187GB-I00, and through AEI under the Severo Ochoa Centres of Excellence Programme 2020--2023 (CEX2019-000920-S).

ARGS acknowledges the supported by FCT through national funds and by FEDER through COMPETE2020 by these grants: UIDP/04434/2020; PTDC/FIS-AST/30389/2017 \& POCI-01-0145-FEDER-030389. ARGS is supported by FCT through the work contract No. 2020.02480.CEECIND/CP1631/CT0001.

We acknowledge that this research was supported in part by the National Science Foundation under Grant No. NSF PHY-1748958.

R.A.G. acknowledges the support of the PLATO CNES grant.


\section*{Data Availability}

The data underlying this article are available in the article and in its online supplementary material.



\bibliographystyle{mnras}
\bibliography{example} 




\appendix

\section{Extra material}
\label{sec:apx1}

\onecolumn

\begin{figure}

	\includegraphics[width=\columnwidth]{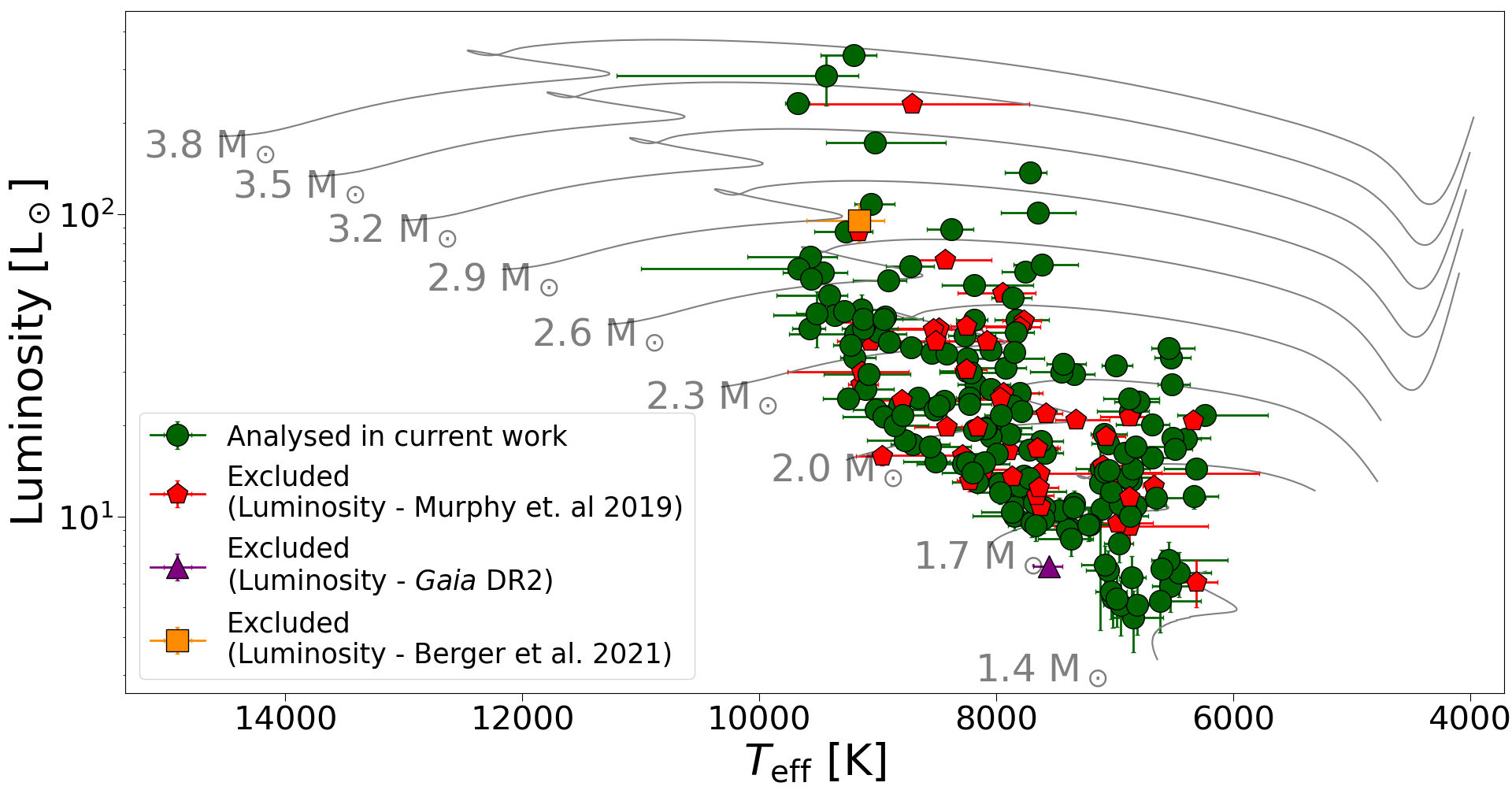}
\caption{HR diagram of the \textit{hump \& spike} stars. Green circles depicts the stars studied in the current work. The source for luminosity and $T_{\rm eff}$ values is described in section \ref{sec:lum_teff}. Red pentagons, orange squares and purple triangles represent the stars excluded as they show signs of binarity (see section \ref{sec:sample_select}). One star that was excluded is not shown in the Figure as no value for luminosity was found in literature: KIC\,5458880 ($T_{\rm eff}$ \textit{Gaia} DR2 $= 8704\,{\rm K} ^{+782} _{-1082}$), was identified as an eclipsing by \citealt{2016AJ....151...68K} and shows a clear binary signal in Kepler data. All $T_{\rm eff}$  values are from \textit{Gaia} DR2. Warszaw-New\,Jersey evolutionary tracks ($Z=0.012$, \citealt{2004A&A...417..751A}) are displayed in the background for guidance only.}
    \label{fig:HRD_all}
\end{figure}

\begin{table}
	\centering
	\caption{Extracted spike parameters and stellar parameters. $f_{\rm rot,}, \sigma_{f_{\rm rot}}$: Spike frequency and associated standard deviation; $\mathrm{A}, \mathrm{\sigma_A}$: Spike amplitude and its uncertainty; 
$T_{\rm eff},\,T_{\rm eff_p}, \,T_{\rm eff_m}$: Effective temperature and upper and lower uncertainties - from \textit{Gaia} DR2; $ L, \,L_{\rm p}, \,L_{\rm m}$: Luminosity and upper and lower uncertainties;
$R, \, R_{\rm p}, \,\rm R_{\rm m} $: Radius value, upper and lower uncertainties;
$\rm 2 \tau_{ACF},\rm \sigma_{\tau_{ACF}}$: Spike lifetime and its uncertainty; No. quarters: number of \textit{Kepler} quarters in which data are available;
Source parameter: first letter indicates the source of the luminosity values, second letter indicates the source of the radius values, m = \citet{2019MNRAS.485.2380M}, i = \citet{2020AJ....159..280B}, g = \textit{Gaia} DR2. The full table is available online; here, the first 10 rows are shown for guidance on content and style.}
 	\setlength{\tabcolsep}{2.4pt} 
	
	\label{tab:spike_param1}
	\begin{tabular}{cccccccccccccccccc} %
	
		\hline
		KIC & $f_{\rm rot}$  & $\sigma_{f_{\rm rot}}$  &A &$\mathrm{\sigma_A}$& $T_{\rm eff}$&$T_{\rm eff_p}$&
		$T_{\rm eff_m}$&$ L$&$L_{\rm p}$&$L_{\rm m}$& $R$&$ R_{\rm p}$&$\rm R_{\rm m} $&$\rm 2 \tau_{ACF}$ &$\rm \sigma_{\tau_{ACF}}$&No.&Sources  \\ 
		
		& [$\mathrm{d^{-1}}$] & [$\mathrm{d^{-1}}$]&[ppm] &[ppm]& [K]&[K]&[K]&[${\rm L}_{\odot}$] &[${\rm L}_{\odot}$]& [${\rm L}_{\odot}$]& $[{\rm R}_{\odot}$]& [${\rm R}_{\odot}$]&[${\rm R}_{\odot}$]&[d]&[d]&quarters&param.\\ 
		
		\hline
        1722916&0.553&0.0013&43.8&0.65&7017&260&98&5.4&1.09&1.09&1.4&0.04&0.09&15&0.1&18&m/g\\
        1873552&1.236&0.0016&5.1&0.51&7826&121&86&11.4&1.05&1.06&1.7&0.04&0.05&25&0.2&18&m/g\\
        2157489&0.737&0.0014&53.0&0.65&7361&140&254&9.1&1.05&1.05&1.7&0.13&0.06&33&0.2&18&m/g\\
        2158190&1.016&0.0006&23.7&0.39&8048&42&219&35.6&1.05&1.05&2.8&0.1&0.09&38&0.1&18&m/i\\
        3002336&1.202&0.0004&76.3&0.71&7115&272&217&14.4&1.06&1.07&2.2&0.14&0.16&20&0.2&15&m/g\\
        3222104&1.27&0.0004&22.8&0.74&8821&231&228&23.3&1.08&1.08&2.2&0.08&0.08&36&0.1&18&m/i\\
        3238627&1.537&0.0024&22.8&0.96&7059&98&140&17.4&1.1&1.1&2.5&0.1&0.07&34&0.2&17&m/g\\
        3240406&1.071&0.0013&5.7&0.35&7865&161&201&13.3&1.06&1.06&1.8&0.1&0.07&30&0.1&18&m/g\\
        3337124&1.108&0.0004&22.7&0.58&7835&61&298&12.8&1.05&1.05&1.8&0.14&0.03&37&0.1&18&m/g\\
        3440710&1.25&0.0021&33.3&1.49&6531&256&154&5.9&1.06&1.06&1.6&0.08&0.12&28&0.1&17&m/g\\

		\hline
	\end{tabular}

\end{table}

\begin{figure}

	\includegraphics[width=\columnwidth, height=400pt]{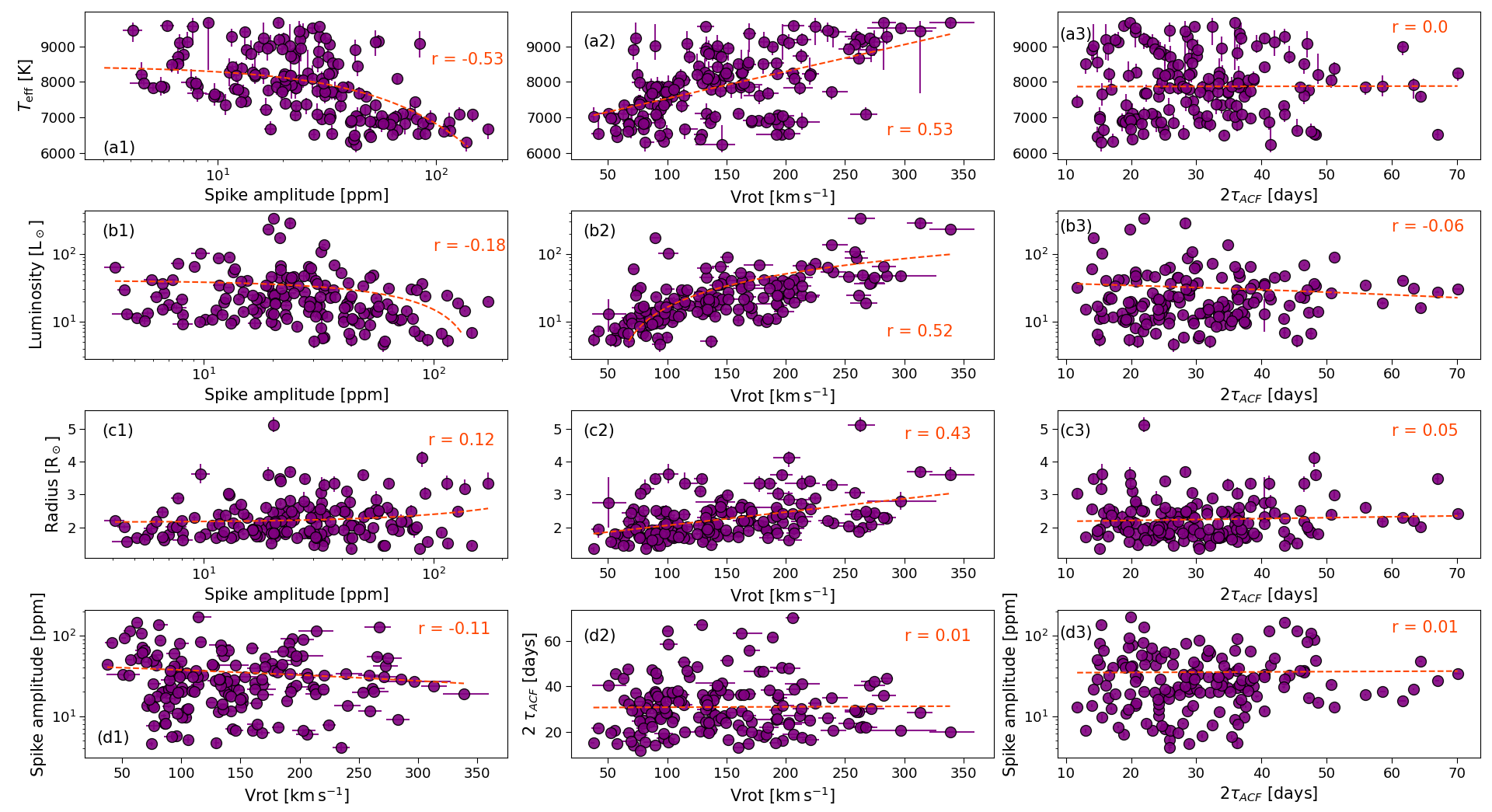}
    \caption{Correlation between stellar parameters and the parameters extracted from our analysis}
    \label{fig:corr_plots}
\end{figure}


\

\bsp	
\label{lastpage}
\end{document}